\documentclass[12pt]{article}
\usepackage{amsmath,amssymb,amsfonts,color,graphicx,cite,color,soul,axodraw}
\input paperdef

\graphicspath{{figs/}}

\evensidemargin \oddsidemargin
\allowdisplaybreaks

\begin{document}
\thispagestyle{empty}

\def\thefootnote{\fnsymbol{footnote}}

\begin{flushright}
\end{flushright}

\vspace{0.5cm}

\begin{center}

{\large\sc {\bf Exploring Quantum Physics at the ILC}}



\vspace{1cm}

{\sc
A.~Freitas$^{1}$%
\footnote{email: afreitas@pitt.edu}%
, K.~Hagiwara$^{2}$%
\footnote{email: kaoru.hagiwara@kek.jp}%
, S.~Heinemeyer$^{3}$%
\footnote{email: Sven.Heinemeyer@cern.ch}%
, P.~Langacker$^{4,5}$%
\footnote{email: pgl@ias.edu}
,\\[.3em] K.~Moenig$^{6}$%
\footnote{email: klaus.moenig@desy.de}%
, M.~Tanabashi$^{7,8}$%
\footnote{email: tanabash@eken.phys.nagoya-u.ac.jp}%
~and G.W.~Wilson$^{9}$%
\footnote{email: gwwilson@ku.edu}%
}

\vspace*{.7cm}

{\sl
$^1$Pittsburgh Particle physics, Astrophysics, and Cosmology
    Center,  Department of Physics \& Astronomy,
    University of Pittsburgh, Pittsburgh, PA 15260, USA

\vspace*{0.1cm}

$^2$KEK Theory Center and Sokendai, Tsukuba 305-0801, Japan

\vspace*{0.1cm}

$^3$Instituto de F\'isica de Cantabria (CSIC-UC), Santander,  Spain

\vspace*{0.1cm}

$^4$Institute for Advanced Study,  Princeton, NJ 08540, USA

\vspace*{0.1cm}

$^5$Department of Physics, Princeton University, Princeton, NJ 08544, USA

\vspace*{0.1cm}

$^6$DESY Zeuthen, Germany

\vspace*{0.1cm}

$^7$Kobayashi-Maskawa Institute for the Origin of Particles and the Universe,
Nagoya University, Nagoya 464-8602, Japan

\vspace*{0.1cm}

$^8$Department of Physics, Nagoya University, Nagoya 464-8602, Japan

\vspace*{0.1cm}

$^9$Department of Physics and Astronomy, University of Kansas, Lawrence, KS 66045, USA

\vspace*{0.1cm}

}

\end{center}

\vspace*{0.1cm}

\begin{abstract}
\noindent
We review the ILC capabilities to explore the electroweak (EW) sector
of the SM at high precision and the prospects of unveiling signals of
BSM physics, either through the presence of new particles in
higher-order corrections or via direct production of extra EW gauge
bosons. This includes electroweak precision observables, global fits to
the SM Higgs boson mass as well as triple and quartic gauge boson couplings.
\end{abstract}

\def\thefootnote{\arabic{footnote}}
\setcounter{page}{0}
\setcounter{footnote}{0}

\newpage


\section{Introduction}
\label{sec:intro}

The Standard Model (SM) cannot be the ultimate fundamental theory of particle
physics. So far, it succeeded in describing direct experimental data
at collider experiments exceptionally well with only a few notable
exceptions, e.g., the left-right ($A_{\rm LR}^e$(SLD)) and
forward-backward ($A_{\rm FB}^b$(LEP)) asymmetry, and the muon magnetic moment
$(g-2)_\mu$.  However, the SM fails to
include gravity, it does not provide cold dark matter, and it has no
solution to the hierarchy problem, i.e.\ it does not have an
explanation for a Higgs-boson mass at the electroweak scale.  On wider
grounds, the SM does not have an explanation for the three generations
of fermions or their huge mass hierarchies.  In order to overcome (at
least some of) the above problems, many new physics models (NPM) have
been proposed and studied, such as
supersymmetric theories, in particular the Minimal Supersymmetric
Standard Model (MSSM), Two Higgs Doublet Models (THDM), Technicolor,
little Higgs models, or models with (large, warped, or universal)
extra spatial dimensions.

If a direct discovery of new BSM particles is out of reach at the LHC
and/or the ILC, precision
measurements of SM observables have proven to be a powerful probe of
NPM via virtual effects of the additional NPM particles.  
In general, precision observables (such as particle masses, mixing
angles, asymmetries etc.)  that can be predicted
within a certain model, including higher order corrections in
perturbation theory, and thus depending sensitively on the other model
parameters, and that can be measured with equally high precision,
constitute a test of the model at the quantum-loop level. Various
models predict different values of the same observable due to their
different particle content and interactions. This permits to
distinguish between, e.~g., the SM and a NPM, via precision
observables. Naturally, this requires a very high precision of both
the experimental results and the theoretical predictions. 

We review the ILC capabilities to explore the electroweak (EW) sector
of the SM at high precision and the prospects of unveiling signals of
BSM physics, either through the presence of new particles in
higher-order corrections or via direct production of extra EW gauge
bosons.  We discuss the experimental and theory uncertainties in the
measurement and calculation of electroweak precision observables
(EWPO), such as the $W$ boson mass and $Z$ pole observables, in
particular the effective weak mixing angle, $\sweff$.
As an example for BSM physics the MSSM is a prominent showcase and will
be used here for illustration.

The recent discovery of a Higgs-like particle at the
LHC has a profound impact on EW precision tests of the SM.  We review
the results of a global EW fit including ILC precision and discuss also
the relevance of a precise top quark mass determination.

We review
the anticipated accuracies for precision measurements of triple and
quartic EW gauge boson couplings. These
observables are of special interest at the ILC, since they have the
potential of accessing energy scales far beyond the direct kinematical
reach of the LHC or the ILC.  Finally, we discuss the ILC
reach for a discovery of extra EW gauge bosons, $Z'$ and $W'$.


\section{The ILC Project Overview}
\label{sec:ILCproject}


Following an intense and successful R\&D phase, 
the ILC has now achieved a state of maturity and readiness, 
culminating recently with the publication of the Technical Design
Report~\cite{TDR}. 
Several important physics goals 
at the TeV energy scale have motivated this effort. 
These include precision measurements of the properties of 
the recently discovered Higgs-like boson, 
including its couplings to fermions and bosons, improving knowledge of 
the top quark to a high level of precision, and the 
search for signals of new physics through the electroweak production 
of new particles and indirectly through precision measurements 
of $W$, $Z$, and two-fermion processes. The ILC experiments will be 
sensitive to new phenomena, such as supersymmetric partners of 
known particles, new heavy gauge bosons, 
extra spatial dimensions, and particles connected with 
strongly-coupled theories of electroweak symmetry breaking~\cite{TDR_Physics}. 
In all of these, the ILC will yield substantial improvements over LHC
measurements and will have a qualitative advantage on signatures that 
have high backgrounds at LHC or are difficult to trigger on. 
Detailed simulations with realistic detector designs show that 
the ILC can reach the precision goals needed~\cite{TDR_DBDs}. 
Just as the LHC experiments are now making more precise measurements 
than were originally predicted (as was also the case with the Tevatron, LEP
and SLC experiments), the ILC experiments will bring qualitatively 
new capabilities and should similarly exceed the performance 
levels based on current simulations when data are in hand.

The requirements of the ILC~\cite{ILC_Parameters_Document} 
include tunability between center-of-mass energies 
of 200 and 500~GeV, with rapid changes in energy over a limited range for 
threshold scans. The luminosity, which must 
exceed $10^{34}$ cm$^{-2}$s$^{-1}$ at 500~GeV, roughly scales 
proportionally with center-of-mass collision energy. Highly polarized
electrons ($> 80 \%$) are specified, with polarized positrons desirable. 
The TDR design~\cite{TDR} has met these specifications. R\&D has achieved the
accelerating gradient goal of 35~MV/m in test stands and 31.5~MV/m in
installed cryomodules with beam loading. Cavity fabrication to these
specifications has been industrialized. The effects of the electron cloud in
the positron damping ring have been studied experimentally, leading to proven
techniques for its mitigation. Fast kickers needed for damping ring beam
injection and ejection have been developed. 
The required small final focus spot size is being demonstrated in a test
facility. The final focus and interaction region, including the detector
push-pull system, has been designed. Two detailed detector designs have been
developed~\cite{TDR_Physics}, with R\&D supporting these designs. Beam tests
with highly granular calorimeters have demonstrated the calorimetry
performance needed by using the particle flow technique.  Similarly, tracking
R\&D has advanced for vertex detection based on thin CMOS monolithic pixel
sensors, outer tracking with low-mass supported silicon microstrips, and  
advanced TPC technologies employing micropattern gas detectors 
or silicon sensors for readout.

Recently, the Japanese government has expressed a desire to host the ILC, and 
international negotiations are underway. In a staged approach, 
beginning at a center-of-mass energy of 250~GeV, a physics program 
would start with precision measurements of the Higgs branching ratios 
and properties. Raising the energy to 500~GeV would move to 
precision measurements of top quark properties well beyond those possible at
the LHC. Measurements of the top coupling to the Higgs and the Higgs self
coupling would begin at 500~GeV.   
Should there be accessible new particles such as supersymmetric 
partners of gauge bosons and leptons, the ILC is the only place where they can
be studied in full detail. If there are multiple Higgs bosons, the ILC would
be needed to measure their branching fractions and the mixing angle
$\tb$.  
Extension of the ILC to 1~TeV is straightforward, with lengthened linac tunnels 
and additional cryomodules, building on the original ILC sources, 
damping rings, final focus and interaction regions, and beam dumps. 


\section{Experimental considerations}
\label{sec:expoverview}

Experiments at $e^+e^-$ colliders play leading roles 
in our current understanding of nature. 
The experimental techniques, conditions and detectors allow the experimenter 
to investigate the science in a direct, largely model-independent 
and above all simple manner. 

The ILC detectors and the ILC facility 
offers the potential to advance particle physics with an unparalleled 
science opportunity. The ILC {\it linear} collider is capable of operating over 
an extensive range of center-of-mass energies colliding point-like particles 
at $\sqrt{s} = 91 - 1000 \gev$ with high luminosity. 
The beam energies are easily tuned to allow threshold scans 
for precision mass measurements. The electron beam can be polarized 
to 80-90\% and the positron beam can also be polarized to 30-60\%. Longitudinal 
polarization is very important~\cite{MoortgatPick:2005cw}. 
Since in the Standard Model the left- and the right-handed electron belong to
different multiplets this gives access to completely different couplings.
It also serves as a method to
enhance/decrease particular processes. An often occurring example is the 
utility of either enhancing or decreasing the contribution of $WW$-fusion and
$WW$ production diagrams. In these respects, polarization is essentially 
a significant increase (up to a factor of 3) in the effective luminosity of
the collider. Physics at an $e^+e^-$ collider enjoys the luxury of
``democratic'' production  
of signal and background leading to very favorable signal to background ratios
even for processes such as those with hadronic final states which can be
buried under many orders of magnitude of QCD background at the LHC.

The ILC experimental outlook is one of a scientific 
facility which will significantly exceed the capabilities of the previous 
generation of experiments such as LEP and SLC. It is in many respects 
targeting much higher performance than is achievable in a hadron collider
environment. A distinct advantage of the ILC linac technology is the time
structure of the colliding bunches. The time between bunches (366~ns), 
the 199~ms quiet time between bunch trains and current technologies lead to a 
data acquisition approach which will greatly benefit the physics. 
There is no need for a hardware-based fast trigger nor concerns on overall
data volume. Furthermore the pulse structure with the possibility of
power-pulsing and the lack of significant radiation hardness constraints,
means that for many subdetector designs the material thicknesses can be
minimized or the technology with the best performance chosen, leading to
better performance and a more hermetic detector. Detector backgrounds are well
understood and not a serious impediment to experimentation. 

These conditions allow the deployment of state-of-the-art pixellated vertex
detectors, large volume thin Silicon tracking, a new generation Time
Projection Chamber (TPC), ultra-high granularity calorimetry designed from the
outset for particle-flow based jet reconstruction, and allows hermetic
coverage to forward angles (around 10~mrad). This results in superb
vertex-tagging of $b$-quarks, $c$-quarks and $\tau$~leptons, exquisite
momentum resolution, jet energy resolution allowing the separation of
hadronically-decaying $W$ and $Z$ bosons, and detection of  
significant missing energy in low visible energy final states. The vision is
to have bubble-chamber like event reconstruction with high multiplicity final
states and essentially no significant background from multiple
interactions. The detectors promise outstanding performance which should be
capitalized on by ensuring that experimental issues like detector stability,
alignment, calibration, magnetic field mapping, 
jet energy scale, momentum scale will be under correspondingly good control. 
 
An extremely important aspect of an $e^+e^-$ collider is the knowledge,
precise measurement and monitoring of the initial-state beam
parameters. Essential quantities to control as well as possible are: the beam
energy, the center-of-mass energy, beam polarizations, luminosity, integrated
luminosity, the luminosity spectrum, the beam energy spread and the
beam-spot. When these are well controlled the physics benefits are
significant. Firstly, a well understood integrated luminosity 
improved to that what was 
achieved at LEP (0.034\% experimental error) allows precision tests of all
absolute cross-section measurements with corresponding theoretical
calculations. This aspect of outstanding absolute normalization of
experimental measurements and the of the dominantly electroweak based theory
calculations, means that even very small deviations can be detected. It also
has great benefits in the direct search for new physics. Secondly, when the
distributions of the initial state four-momenta are well measured, one can in
many circumstances take advantage of the kinematic constraints of energy and
momentum conservation. Both of these major advantages are foreseen for the
ILC, and significant preparatory work has already established the feasibility
of quality knowledge of the beam parameters. See for example~\cite{List} for
related work on applying kinematic constraints. 

One of the new features of ILC compared to LEP and SLC is that the highly
focused beams (several nm vertical beam size) lose on average a few per-cent
of their energy through the emission of beamstrahlung photons as a result of
the beam-beam interaction. The overall effect for ILC is small on the scale of
initial state radiation but must be measured directly from collision events
such as Bhabha scattering events~\cite{Moenig-and-Sailer-Diplomarbeit}.

When electron and positron polarization is available the beam polarization can
be measured from data in a model independent way and polarimeters are needed
only for small corrections. Also the beam energy can be calibrated relative to
the Z-boson mass which is known from LEP with a precision of 2.1~MeV. To go
beyond this precision an absolute beam energy measurement on the $10^{-5}$
level is needed which seems difficult.

The ILC will run in several distinct phases. Initially the ILC will run at a
center of mass energy around 250~GeV for a precise measurement of the Higgs
couplings and then gradually increase its energy via the top-pair threshold at
350~GeV to its maximum value. From the beginning runs at the top of the
Z-resonance will be needed to calibrate the energy scale of the detector.  At
a later stage a long run at the Z pole is foreseen to collect about $10^9$ Z
decays for a measurement of the effective weak mixing angle with ultimate
precision. This run also requires a scan of the resonance which might also be
used to improve the knowledge of the Z width. If it turns out to be
interesting the W-mass measurement can be improved with a dedicated scan of
the W-pair production threshold around 160~GeV.
  
Some of the relevant beam parameters are shown in \refta{tab:ILCpara}
for the default TDR parameter sets used in 
the TDR full simulation studies.
It should be noted that the momentum spread at 200~GeV is very similar to LEP2.
     
%

\begin{table}[ht!]
\renewcommand{\arraystretch}{1.3}
\centering
\begin{tabular}{|c||c|c|c|c|}
\hline
$\sqrt{s}$ & $\cL [10^{34}]$ & dE [\%] & (dp/p)(+) [\%] & (dp/p)(-) [\%] \\
\hline\hline
200 & 0.56 & 0.65 & 0.190 & 0.206 \\ \hline
250 & 0.75 & 0.97 & 0.152 & 0.190 \\ \hline
350 & 1.0 & 1.9 & 0.100 & 0.158 \\ \hline
500 & 1.8/3.6 & 4.5 & 0.070 & 0.124 \\ \hline
1000 & 4.9 & 10.5 & 0.047 & 0.085 \\
\hline\hline
\end{tabular}
\caption{Relevant beam parameters for the ILC as a function of $\sqrt{s}$.
Given are the luminosity, the average energy loss from beamstrahlung, 
average center-of-mass energy spread from momentum spread.
}
\label{tab:ILCpara}
\renewcommand{\arraystretch}{1}
\end{table}


\section{Electroweak precision observables}
\label{sec:ewpo}


\subsection{The \boldmath{$W$} boson mass}
\label{sec:mw}

The mass of the $W$ boson is a fundamental parameter of the
electroweak theory and a crucial input to electroweak precision tests.
The present world average for the W-boson mass~\cite{Group:2012gb},
\begin{align}
\MW^{\rm exp} &= 80.385 \pm 0.015 \gev~,
\label{mwexp}
\end{align}
is dominated by the results from the Tevatron, where the $W$ boson
mass has been measured in Drell--Yan-like single-$W$-boson production.
At LEP2, the $W$-boson mass had been measured in $W$-pair production
with an error of $33\mev$ from direct reconstruction and $\sim200\mev$
from the cross section at threshold \cite{Alcaraz:2006mx}.  

The three most promising approaches to measuring the $W$~mass at the ILC
are: 
\begin{itemize}
\item Polarized threshold scan of 
the $W^+W^-$ cross-section as discussed in~\cite{Wilson_Sitges}.
\item Kinematically-constrained reconstruction of $W^+W^-$ using 
constraints from four-momentum conservation and optionally mass-equality 
as was done at LEP2.
\item Direct measurement of the hadronic mass. This can be applied 
particularly to single-$W$ events decaying hadronically 
or to the hadronic system in semi-leptonic $W^+W^-$ events.
\end{itemize}

Each method can plausibly measure $\MW$ to an experimental precision in
the $5-6 \mev$ range. 
The three methods are largely uncorrelated. If all three methods do live 
up to their promise, one can target an overall uncertainty 
on $\MW$ in the range of $3-4 \mev$.

The anticipated experimental accuracy has to be matched with a
theoretical uncertainty at the same level of accuracy. This is
particularly challenging for the $WW$ threshold scan, where a full
diagrammatic calculation of $e^+e^- \to 4f$ and leading two-loop
corrections are required, see
\citeres{Denner:2005es,Denner:2005fg} and references therein.
All building blocks for a sufficiently precise prediction
of the $W$-pair production cross section in the threshold region are
available. They require the combination of the NLO calculation of the
full four fermion cross section with the (parametrically) dominant
NNLO corrections, which are calculated within the EFT. 
For the precise determination of the
cross section at energies above $500\gev$ the leading
two-loop (Sudakov) corrections should be included in addition
to the full NLO corrections.
Combining the theoretical uncertainties with the anticated precision from a
threshold scan (see the previous subsection) a total uncertainty of $6 \mev$
can be estimated (see also \citere{Baur:2001yp}).
For the overall future experimental uncertainty we arrive at
\begin{align}
\de\MW^{\rm exp,ILC} = 5-6 \mev~.
\label{mwexpfut}
\end{align}

The currently most accurate theoretical prediction of $\MW$ in the SM is based
on a full two-loop calculation, supplemented with leading corrections at the
three- and four-loop level, entering via the $\rho$-parameter, see
\citere{Heinemeyer:2004gx} for a review (and references therein). The
total intrinsic uncertainty from unknown higher-order corrections 
has been estimated to~\cite{deMWSMtheo}
\begin{align}
\de\MW^{\rm SM,theo} = 4 \mev~.
\end{align}
This error mainly stems from missing $\order{\al^2\als}$, $\order{N_f^3\al^3}$
and $\order{N_f^2\al^3}$ contributions, where $N_f^n$ denotes diagrams with
$n$ closed fermion loops. These are expected to be calculable in the
forseeable future using numerical methods or asymptotic expansions, leading to
a remaining intrinsic uncertainty of
\begin{align}
\de\MW^{\rm SM,theo,fut} &\approx 1\mev~.
\end{align}
Within the MSSM, due to the additional missing higher-order corrections
the current intrinsic theory uncertainty increases to~\cite{Heinemeyer:2006px}
\begin{align}
\de\MW^{\rm MSSM,theo} = 5-10 \mev~,
\end{align}
depending on the masses of the SUSY particles. 
In the future this could be reduced to 
\begin{align}
\de\MW^{\rm MSSM,theo,fut} = 3-5 \mev~.
\end{align}
The main SM parametric uncertainties are introduced by $\mt$,
$\Dalhad$ and $\MZ$. Using today's values $\de\mt = 0.9\gev$,
$\de(\Dalhad) = 10^{-4}$ and $\de\MZ = 2.1 \mev$, 
yields
\begin{align}
\de\MW^{{\rm para,}\mt} = 5.5 \mev, \quad
\de\MW^{{\rm para,}\Dalhad} = 2 \mev, \quad
\de\MW^{{\rm para,}\MZ} = 2.6 \mev~,
\label{mwpara}
\end{align}
dominating the intrinsic theory uncertainties. They are reduced,
however, if one assumes the ILC accuracy for $\mt$ as given by
\refeq{mtexpfut}, and an improvement to 
$\de(\Dalhad) = 5 \times 10^{-5}$ by future low energy $e^+e^-$
measurements, 
\begin{align}
\De\MW^{{\rm para,fut,}\mt} = 1 \mev, \quad
\De\MW^{{\rm para,fut,}\Dalhad} = 1 \mev~,
\label{mwparafut}
\end{align}
where no improvement in the measurement of $\MZ$ is current
foreseable. Thus $\De\MW^{{\rm para,}\MZ}$ could dominate the future
theoretical uncertainties. However, they would still stay below the
anticipated ILC experimental accuracy as given in \refeq{mwexpfut}. 

\begin{figure}[htb!]
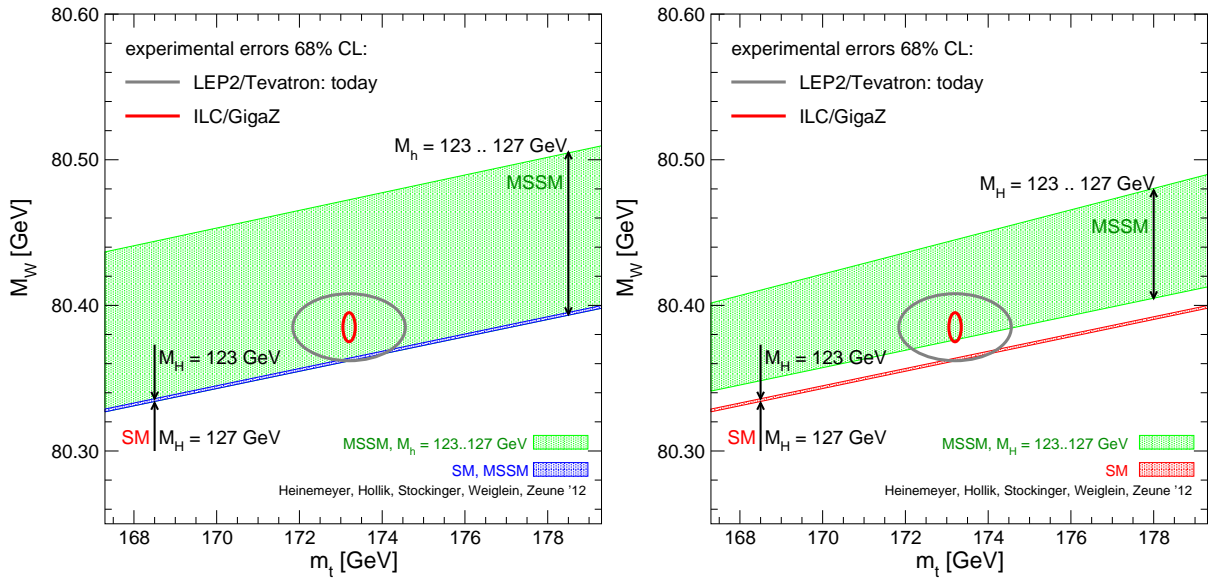

\centering
\includegraphics[width=0.48\columnwidth]
        {MWmt12LEPTevILCGigaZ.123-127-h.1732.80385.cl.eps}
\includegraphics[width=0.48\columnwidth]
        {MWmt12LEPTevILCGigaZ.123-127-H.1732.80385.cl.eps}
\caption{Prediction for $\MW$ as a function of $\mt$.
The left plot shows the $\MW$ prediction assuming the light
$\cp$-even Higgs $h$ in the region $125.0 \pm 2 \gev$. 
The blue band indicates the overlap region of the SM and the MSSM
with $\MHSM = 125.0 \pm 2 \gev$.
The right plot shows the $\MW$ prediction assuming the heavy $\cp$-even
Higgs $H$ in the region $125.0 \pm 2 \gev$.  
The red band indicates the SM region 
with $\MHSM = 125.0 \pm 2 \gev$.
The gray ellipse indicates the current experimental uncertainty, whereas the
red ellipse shows the anticipated future ILC/GigaZ precision.
} 
\label{fig:mtmwmh125}
\end{figure}

\bigskip
An example that demonstrates the power of the ILC precision in $\MW$,
taken together with the top quark mass (see \refse{sec:mt}), is shown in
\reffi{fig:mtmwmh125}. 
The evaluation of $\MW$ includes the full one-loop result and all known higher
order corrections of SM- and SUSY-type, for details
see~\cite{MWlisa,Heinemeyer:2006px} and references therein. 
In the left plot the green region indicated the MSSM $\MW$ prediction
(as obtained from a 15-dim.\ parameter scan~\cite{MWlisa}), 
assuming the light $\cp$-even Higgs $h$ in the region $125.0 \pm 2 \gev$. 
The blue band indicates the overlap region of the SM and the MSSM
with $\MHSM = 125.0 \pm 2 \gev$.
The right plot shows the $\MW$ prediction assuming the heavy $\cp$-even
Higgs $H$ in the region $125.0 \pm 2 \gev$.  
The red band indicates the SM region
with $\MHSM = 125.0 \pm 2 \gev$.
The gray ellipse indicates the current experimental uncertainty,
see \refeqs{mwexp}, (\ref{mtexp}), whereas the
red ellipse shows the anticipated future ILC/GigaZ precision, assuming 
an uncertainty of $7 \mev$ for $\MW$ and $0.1 \gev$ for $\mt$.
While at the current level of precision SUSY might be considered as slightly
favored over the SM by the $\MW$-$\mt$ measurement, no clear conclusion can be
drawn. The small red ellipses, on the other hand, indicate the discrimination
power of the future ILC/GigaZ measurements. With the improved precision a
small part of the MSSM parameter space could be singled out. The
comparison of the SM and MSSM predictions with the ILC/GigaZ precision could
rule out either of models.

\subsection{The \boldmath{$Z$} boson observables}
\label{sec:zobs}

\newcommand{\vef}{g_{V_e}}
\newcommand{\aef}{g_{A_e}}
\newcommand{\ppl}{{\cal P}_{e^+}}
\newcommand{\pmi}{{\cal P}_{e^-}}
\newcommand{\ppm}{{\cal P}_{e^\pm}}
\newcommand{\peff}{{\cal P}_{\rm{eff}}}
\newcommand{\ALRe}{A_{\rm LR}^e}

Other important EWPOs are the various observables related to the $Z$
boson, measured in four-fermion processes, 
$e^+ e^- \to \gamma,Z \to f \bar f$, at the $Z$~boson pole. 
Besides the improvements in $\sweff$ (which will be discussed below in
detail) and $\MW$, GigaZ has the potential 
to determine the total $Z$~width within $\de\Gamma_Z = \pm 1$~MeV; the ratio 
of hadronic to leptonic partial $Z$~widths with a relative uncertainty of 
$\de R_l/R_l = \pm 0.05 \%$; the ratio of the $b\bar{b}$ to the hadronic 
partial widths with a precision of $\de R_b = \pm 1.4 \times 10^{-4}$; and 
to improve the $b$ quark asymmetry parameter $A_b$ to a precision of
$\pm 1 \times 10^{-3}$~\cite{gigaz,gigazsitges}. 

A special role is played by the effective weak
leptonic mixing angle, $\sweff$, which can be determined via various
measurements, in particular via the forward-backward (FB) asymmetry of
$b$~quarks, $A_{\rm FB}^b$, and via the left-right (LR) asymmetry of
electrons, $\ALRe$.
The current experimental uncertainty is given by \cite{lepz}
\begin{align}
\sweff^{\rm \,exp} = 0.23153 \pm 0.00016~,
\label{sweffexp}
\end{align}
mainly driven by
\begin{align}
\label{afb}
A_{\rm FB}^b({\rm LEP}) &: \sweff^{\rm \,exp,LEP} = 0.23221 \pm 0.00029~, \\[.1em]
\label{alr}
\ALRe({\rm SLD}) &: \sweff^{\rm \,exp,SLD} = 0.23098 \pm 0.00026~.
\end{align}
At the ILC $\sweff$ can be measured running at the $Z$-mass (i.e.\ at
GigaZ), using the left-right asymmetry~\cite{gigaz}.
With at least the electron beam polarised with a polarisation of ${\cal P}$,
$\sweff$ can be obtained via
\begin{align}
\label{eq:alrdef}
\ALRe &=  \frac{1}{{\cal P}}\frac{\si_L-\si_R}{\si_L+\si_R}
     = \frac{2 \vef \aef}{\vef^2 +\aef^2} \\
  {\vef}/{\aef} &=  1 - 4 \sweff 
\end{align}
independent of the final state. $\vef$ and $\aef$ denote the vector and
axial-vector couplings of the $Z$~boson to electrons.
With $10^9$ $Z$~bosons, an electron polarisation of 80\% and no positron
polarisation the statistical error is $\De \ALRe = 4 \times 10^{-5}$.
The error from the polarisation measurement is
$\De \ALRe/\ALRe = \Delta {\cal P}/{\cal P}$.
With electron polarisation only and 
$\De {\cal P}/{\cal P} = 0.5\%$  one has
$\De \ALRe = 8 \times 10^{-4}$, 
much larger than the statistical precision. 
If  positron polarisation is also available ${\cal P}$ in equation
(\ref{eq:alrdef}) has to be replaced by 
$\peff \, = \, \frac{\ppl+\pmi}{1+\ppl\pmi}$. 
For $\pmi(\ppl) = 80\%(60\%)$ 
the error in $\peff$ is a factor of three to four smaller than the error on
$\ppl,\, \pmi$ depending on the correlation between the two
measurements. If one takes, however, data on all four polarisation
combinations the left-right asymmetry can be extracted without absolute
polarimetry \cite{Blondel:1987wr} and basically without increasing the error if
the positron polarisation is larger than 50\%. Polarimetry, however, is
still needed for relative measurements like the difference of absolute
values of the positive and the negative helicity states.
Assuming conservatively $\De \ALRe = 10^{-4}$ leads to 
\begin{align}
\de\sweff^{\rm \,exp,ILC} = 0.000013~,
\label{sweffexpfut}
\end{align} 
more than a factor 10 better than the LEP/SLD result.

Within the SM, a full SM two-loop calculation for $\sweff$ is
available, which is supplemented by the same type of three- and
four-loop corrections as for $\MW$. This yields an intrinsic uncertainty
of~\cite{sw2effSM,sw2effSM2}
\begin{align}
\de\sweff^{\rm SM,theo} = 4.5 \times 10^{-5}~,
\end{align}
which is mainly due to missing $\order{\al^2\als}$, $\order{N_f^3\al^3}$ and
$\order{N_f^2\al^3}$ contributions. Assuming that these corrections will be
calculated in the future, this error can be reduced to
\begin{align}
\de\sweff^{\rm SM,theo,fut} \approx 1.5 \times 10^{-5}~.
\end{align}
Within the MSSM the intrinsic uncertainty increases relative to the SM due to
the additional unknown higher-order corrections to~\cite{Heinemeyer:2007bw}
\begin{align}
\de\sweff^{\rm MSSM,theo} = (5-7) \times 10^{-5}~,
\end{align}
depending on the relevant SUSY mass scales. In the future one can expect
a reduction to
\begin{align}
\de\sweff^{\rm MSSM,theo,fut} = (2.5 - 3.5) \times 10^{-5}~.
\end{align}
The current parametric uncertainties (see the previous subsection) read
\begin{align}
\de\sweff^{{\rm para,}\mt} = 7 \times 10^{-5}, \quad
\de\sweff^{{\rm para,}\Dalhad} = 3.6 \times 10^{-5}, \quad
\de\sweff^{{\rm para,}\MZ} = 1.4 \times 10^{-5}~,
\label{sw2effpara}
\end{align}
to be improved in the future (see the previous subsection) to
\begin{align}
\de\sweff^{{\rm para,fut,}\mt} = 0.4 \times 10^{-5}, \quad
\de\sweff^{{\rm para,fut,}\Dalhad} = 1.8 \times 10^{-5}~.
\label{sw2effparafut}
\end{align}
The parametric uncertainties induced by $\MZ$ and in particular
$\Dalhad$, even assuming the future precision, are at or even slightly
above the GigaZ precision of $\sweff$. Consequently, it will be very
challenging to fully exploit the GigaZ precision.

\begin{figure}[htb!]
\begin{center}
\includegraphics[width=0.88\textwidth]{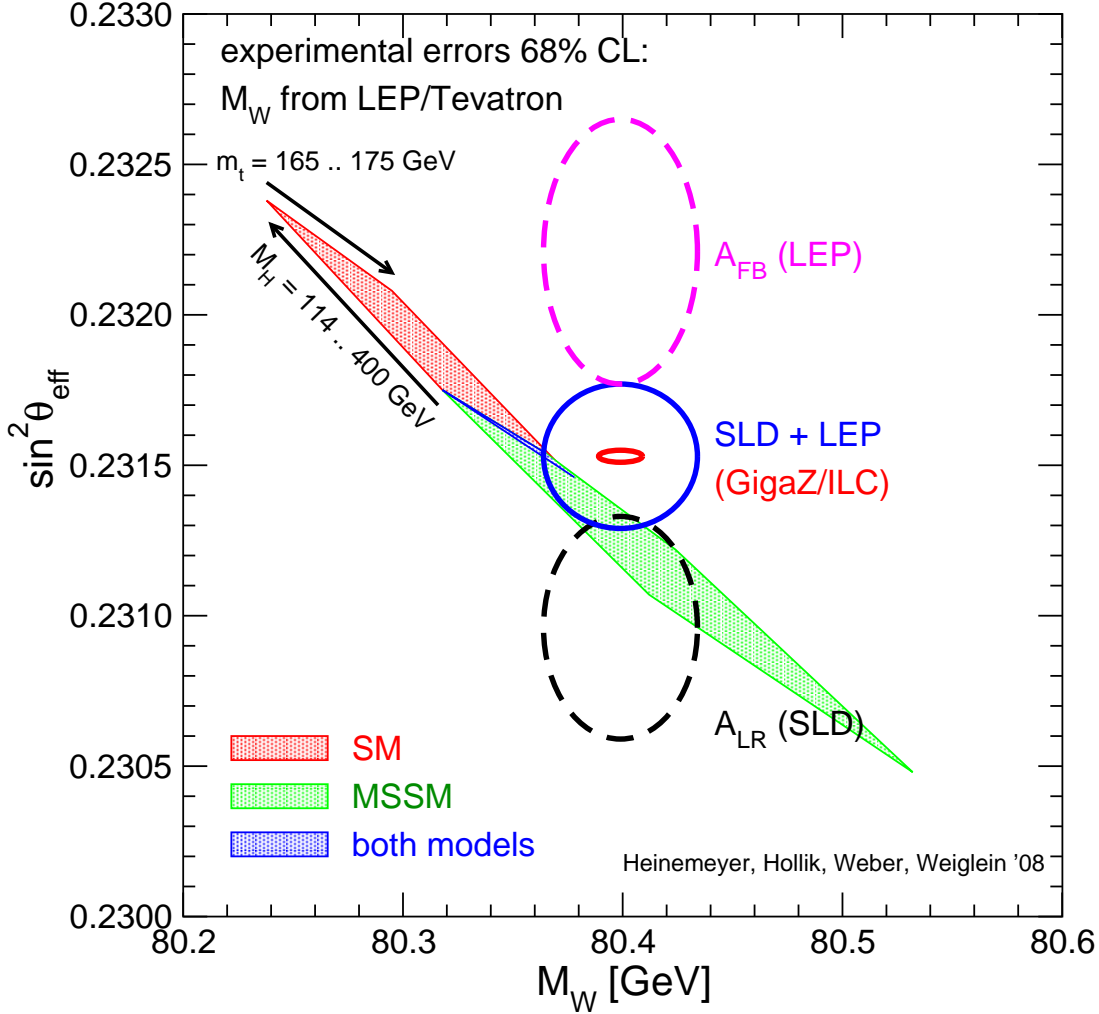}
\end{center}
\vspace{-1em}
\caption{MSSM parameter scan for $\MW$ and $\sweff$ over 
  ranges similar to \reffi{fig:mtmwmh125} with 
  $\mt = 165 \ldots 175 \gev$. Todays 68\%~C.L.\ ellipses (from 
  $A_{\rm FB}^b({\rm LEP})$, $A_{\rm LR}^e({\rm SLD})$ and the world
  average) are shown 
  as well as the anticipated GigaZ/ILC precisions, drawn around today's
  central value.}  
\label{fig:sw2efftheo-Scans2} 
\end{figure}

\bigskip
In \reffi{fig:sw2efftheo-Scans2} we
compare the SM and the MSSM predictions for $\MW$ and $\sweff$
as obtained from scatter data similar to the one used in
\reffi{fig:mtmwmh125}. 
The predictions within the two models 
give rise to two bands in the $\MW$--$\sweff$ plane with only a relatively 
small overlap region (indicated by a dark-shaded (blue) area).
The parameter region shown in the SM (the medium-shaded (red)
and dark-shaded (blue) bands) arises from varying the mass of the SM
Higgs boson, from $\MHSM = 114\gev$, the old LEP exclusion
bound~\cite{Barate:2003sz} 
(lower edge of the dark-shaded (blue) area), to $400 \gev$ (upper edge of the
medium-shaded (red) area), and from varying $\mt$ in the range of 
$\mt = 165 \ldots 175 \gev$. The value of $\MHSM \sim 125.5 \gev$
corresponds roughly to the dark-shaded (blue) strip.
The light shaded (green) and the
dark-shaded (blue) areas indicate allowed regions for the unconstrained
MSSM, where no restriction on the light $\cp$-even Higgs mass has been
applied. 
Including a Higgs mass measurement into the MSSM scan would cut away
small part at the lower edge of the light shaded (green) area.

The 68\%~C.L.\ experimental results
for $\MW$ and $\sweff$ are indicated in the plot. 
The center ellipse corresponds to the current world average given
in \refeq{sweffexp}. 
Also shown are the error ellipses corresponding to the two individual most
precise measurements of $\sweff$ , based on 
$A_{\rm LR}^e$ by SLD and $A_{\rm FB}^b$ by LEP, corresponding to
\refeqs{afb}, (\ref{alr}).
The first (second) value prefers a 
value of $\MHSM \sim 32 (437) \gev$~\cite{gruenewaldpriv}. 
The two measurements differ by more than $3\,\si$.
The averaged value of $\sweff$ , as given in \refeq{sweffexp},
prefers $\MHSM \sim 110 \gev$~\cite{gruenewaldpriv}. 
The anticipated improvement with the ILC/GigaZ measurements, indicated as
small ellipse, is shown around the current experimental central data.
One can see that the current averaged value is compatible with the SM
with $\MHSM \sim 125.5 \gev$
and with the MSSM. The value of $\sweff$ obtained from $\ALRe$(SLD)
clearly favors the MSSM over the SM.
On the other hand, the value of $\sweff$ obtained from $A^b_{\rm FB}$(LEP) 
together with the $\MW$ data from LEP and the Tevatron would correspond to an
experimentally preferred region that deviates from the predictions of both
models. 
This unsatisfactory solution can only be resolved by new measurements, where
the  $Z$~factory, i.e.\ the GigaZ option would be an ideal solution. 
Thus, the unclear experimental situation regarding the two single
most precise measurements entering the combined value for $\sweff$
has a significant impact on the constraints that can be obtained from this
precision observable on possible New Physics scenarios. Measurements at a new 
$e^+e^-$ $Z$~factory, which could be realized in particular with the GigaZ
option of the ILC, would be needed to resolve this issue.
As indicated by the solid light shaded (red) ellipse, the anticipated
ILC/GigaZ precision of the combined $\MW$--$\sweff$ measurement could
put severe constraints on each of the models and resolve the discrepancy
between the $A_{\rm FB}^b$(LEP) and $\ALRe$(SLD) measurements. 

\bigskip
Besides the leptonic effective weak mixing angle, similar effective weak mixing
angles, $\sin^2{\theta^f}_{\mathrm{eff}}$, can be defined for the interaction
of the $Z$ boson with other fermion 
flavors $f$, $f\neq \ell$. The SM predictions for these quantities have been
computed including two-loop corrections with at least one close fermion loop
(i.e.\ the ``bosonic'' electroweak two-loop contributions without closed
fermion loops are not available yet), as well as leading three-loop
corrections \cite{sw2effSM2,swbb}. 
The remaining intrinsinc theoretical uncertainty is
\begin{align}
\de\sin^2{\theta^{f \neq \ell}}_{\mathrm{eff}}^{\rm SM,theo} 
\approx 5 \times 10^{-5}~,
\end{align}
which will be sufficient for the forseeable future since the experimental
precision for $\sin^2{\theta^b}_{\mathrm{eff}}$,
$\sin^2{\theta^c}_{\mathrm{eff}}$, etc.\ is more than an order of magnitude less
than for the leptonic weak mixing angle~\cite{lepz}.

\begin{figure}[htb!]
\begin{center}
\includegraphics[width=0.6\textwidth]{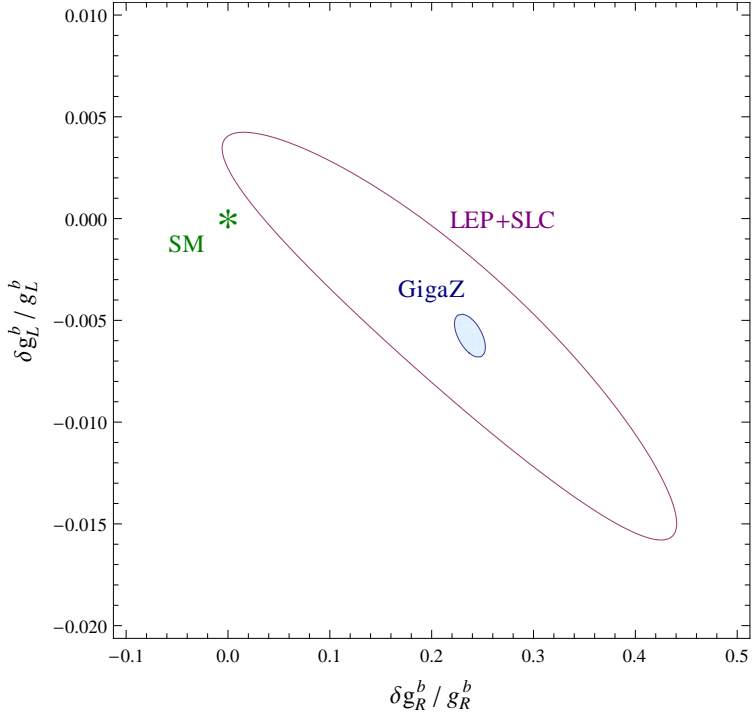}
\end{center}
\vspace{-1.5em}
\caption{95\%~C.L.\ level regions for the left- and right-handed
$Zb\bar{b}$ couplings based on current experimental results from LEP/SLC and
projected precision for GigaZ/ILC (assuming the same central values). The
regions reflect experimental errors only. $\delta
g^b_{L,R}=0$ corresponds to the SM prediction (green star).}
\label{fig:bcoupl}
\end{figure}

Besides asymmetry observables, additional constraints can be obtained from
(partial) $Z$~boson decay widths. The two most relevant quantities in this
context are the total decay width $\Gamma_Z$, which is determined from the
lineshape of the 
cross section $\sigma_{e^+e^- \to f\bar{f}}(s)$, and the ratio
$R_b \equiv \Gamma_{Z\to b\bar{b}}/\Gamma_{Z\to {\rm hadrons}}$. In particular,
$R_b$ is sensitive to new physics in the third generation of fermions, which is
not directly probed by $\MW$ and $\sweff$~\cite{bnew}.
The current experimental result for $R_b$ is \cite{lepz}
\begin{align}
R_b^{\rm \,exp} = 0.21629 \pm 0.00066~.
\label{rbexp}
\end{align}
This value differs from the SM prediction by about two standard deviations,
which can be interpreted in terms of shifted couplings of the $Z$-boson to
left- and right-handed bottom quarks, $g^b_{L,R}$ \cite{bnew}, see
\reffi{fig:bcoupl}.

Due to higher statistics, the ILC running in the GigaZ mode
will be able to reduce the experimental error substantially to \cite{gigaz}
\begin{align}
\de R_b^{\rm \,exp,ILC} = 0.00015~,
\label{rbexpfut}
\end{align}
which, together with improved asymmetry measurements, will help to clearly
identify or rule out a possible new physics effect in the $Zb\bar{b}$
couplings, see Fig.~\ref{fig:bcoupl}.

The currently most precise SM prediction includes fermionic two-loop
corrections (i.e.\ two-loop diagrams with at least one closed fermion loop)
and leading three-loop terms~\cite{rb}, with an intrinsic uncertainty of
\begin{align}
\de R_b^{\rm SM,theo} \approx 2 \times 10^{-4}~.
\end{align}
The leading unkown contributions are \order{\al\als^2} and
\order{\al^2\als} terms. In contrast to $\sweff$, the vertex corrections of
this order for quark final states involve more complex diagram topologies, so
that only an approximate calculation in terms of a large-$\mt$ expansion may
be feasible in the near future. This would reduce the theory uncertainty to
\begin{align}
\de R_b^{\rm SM,theo,fut} = (0.5 - 1) \times 10^{-4}~.
\end{align}
The current parametric uncertainties of $R_b$ are small:
\begin{align}
\de R_b^{{\rm para,}\mt} = 3.5 \times 10^{-5}, \quad
\de R_b^{{\rm para,}\Dalhad} = 1.2 \times 10^{-6}, \quad
\de R_b^{{\rm para,}\MZ} = 1.4 \times 10^{-6}~,
\label{rbpara}
\end{align}
and thus they do not pose any limit to future improvements.

The total $Z$ width, $\Gamma_Z$, has been measured with high precision at LEP
\cite{lepz},
\begin{align}
\Gamma_Z^{\rm \,exp} = 2.4952 \pm 0.0023 \gev~,
\label{gzexp}
\end{align}
which is mainly limited by the calibration and temporal fluctations 
of the center-of-mass energy \cite{lepz}. Therefore, the experimental
precision is expected to improve only moderately at the GigaZ run of the
ILC. A reasonable estimate~\cite{gigaz} gives 
\begin{align}
\de \Gamma_Z^{\rm \,exp,ILC} = 0.001 \gev~.
\label{gzexpfut}
\end{align} 
For the calculation of $\Gamma_Z$ in the SM, only an approximate result for
the electroweak two-loop corrections in the limit of large $\mt$ is known
\cite{gzmt}. The remaining \order{N_f\alpha^2} may be relatively large, as
turned out to be the case for $R_b$~\cite{rb}. Assuming the same relative
size of the these corrections as for $R_b$, this leads to a current intrinsic
uncertainty of a few MeV,
which is by far dominant compared to missing three-loop contributions.
However, the \order{N_f\alpha^2} correction can be computed with existing
methods without conceptual difficulties. The remaining intrinsic uncertainty is
estimated to be
\begin{align}
\de \Gamma_Z^{\rm SM,theo,fut} < 1\mev~.
\end{align}


\subsection{The top quark mass}
\label{sec:mt}

The mass of the top quark, $\mt$, is a fundamental parameter of the 
electroweak theory. It is by far the heaviest of all quark masses and
it is also larger than the masses of all other known fundamental
particles.The large value of $\mt$ gives rise to a large coupling
between the to top 
quark and the Higgs boson and is furthermore important for flavor
physics. It could therefore provide a window to new physics. 
The top-quark mass also plays an important role in electroweak precision
physics, as a consequence in particular of non-decoupling effects being
proportional to powers of $\mt$. A precise knowledge of $\mt$ is
therefore indispensable in order to have sensitivity to possible effects
of new physics in electroweak precision tests, see \refeqs{mwpara},
(\ref{mwparafut}), (\ref{sw2effpara}), (\ref{sw2effparafut}). 

The current world average for the top-quark mass from the measurement
at the Tevatron is~\cite{Lancaster:2011wr},
\begin{align}
\mt^{\rm exp} &= 173.2 \pm 0.9 \gev~.
\label{mtexp}
\end{align}
The prospective accuracy at the LHC is 
$\de\mt^{\rm exp} \approx 1 \gev$,
while at the ILC a very precise determination of $\mt$ with an accuracy
of 
\begin{align}
\de\mt^{\rm exp,ILC} = 0.1 \gev
\label{mtexpfut}
\end{align}
will be possible. 
This uncertainty contains both the experimental error of the mass parameter
extracted from the $t \bar t$ threshold measurements at the ILC and 
the envisaged theoretical uncertainty from its transition into a suitable
short-distance mass (like the \msbar\ mass).


The relevance of the $\mt$ precision as parametric uncertainty has been
discussed for the $W$~boson mass, $\MW$, in \refse{sec:mw}, and
for the effective leptonic weak mixing angle, $\sweff$, in
\refse{sec:zobs}. 

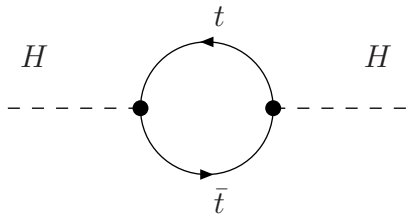
\begin{figure}[htb!]
\BC
\setlength{\unitlength}{1pt}
\begin{picture}(170, 70)
\DashLine(10,40)(60,40){5}
\Text(20,60)[]{$H$}
\Vertex(60,40){3}
\ArrowArc(85,40)(25,0,180)
\ArrowArc(85,40)(25,180,360)
\Text(90,75)[]{$t$}
\Text(90,5)[]{$\bar t$}
\Vertex(110,40){3}
\DashLine(110,40)(160,40){5}
\Text(150,60)[]{$H$}
\end{picture}
\EC
\vspace{-2em}
\caption{Loop contribution of the top quark to the Higgs boson mass.}
\label{fig:mhiggs}
\end{figure}

Because of its large mass, the top quark is expected to have a large
Yukawa coupling to Higgs bosons, being proportional to $\mt$.
In each model where the Higgs boson mass is not a free
parameter but predicted in terms of the the other model parameters
(as e.g.\ in the MSSM), the diagram in \reffi{fig:mhiggs} contributes
to the Higgs mass. This diagram gives rise to a leading $\mt$
contribution of the form
\BE
\De\MH^2 \sim \gf \; N_C \; C \; \mt^4~,
\end{equation}
where $\gf$ is the Fermi constant, $N_C$ is the color factor, and the
coefficient $C$ depends on the specific model. Thus the experimental
error of $\mt$ necessarily leads to a parametric error in the Higgs
boson mass evaluation. 

Taking the MSSM as a specific example 
(including also the scalar top contributions and the appropriate
renormalization) $N_C \, C$ is given for the light $\cp$-even Higgs
boson mass in leading logarithmic approximation by  
\BE
N_C \, C = \frac{3}{\sqrt{2}\,\pi^2\,\SQb} \; 
\log \KL \frac{\mste\mstz}{\mt^2} \KR~.
\end{equation}
Here $m_{\tilde t_{1,2}}$ denote the two masses of the scalar tops.
The current
precision of $\de\mt \sim 1 \gev$ leads to an uncertainty of 
$\sim 2.5\%$ in the prediction of $\MH$, while the ILC will yield a
precision of $\sim 0.2\%$.  
These uncertainties have to be compared with the anticipated precision of
the future Higgs boson mass measurements. With a precision of 
$\de\MH^{\rm exp,LHC} \approx 0.2 \gev$ the relative
precision is at the level of $\sim 0.2\%$. It is apparent that only the
ILC precision of $\mt$ will yield a parametric error small enough to
allow a precise comparison of the Higgs boson mass prediction and its
experimental value (keeping also in mind the intrinsic theoretical
uncertainties on $\MH$, see, e.g., \citere{Degrassi:2002fi} for the case
of the MSSM).


\subsection{The strong coupling constant}
\label{sec:alphas}

The GigaZ run can offer complementary information on $\als$. It was
shown~\cite{Erler:2000jg} that a measurement of 
$R_l := \Ga(Z \to \mbox{hadrons})/\Ga(Z \to \mbox{leptons})$ 
down to $\de R_l/R_l = \pm 0.05\%$
at GigaZ would provide a clean determination of $\als$ with an
uncertainty of 
\begin{align}
\de\als^{\rm exp,ILC} \approx 0.001~.
\label{alsexpfut}
\end{align}
Since $\als$ enters (at least at the two-loop level) the radiative
corrections to precision observables, it is important to control $\als$
effects to avoid confusion with other parametric uncertainties.


\subsection{The Higgs boson mass}
\label{sec:mh}

The observation of a new particle compatible with a Higgs boson of mass
$\sim 125 \gev$ is a major breakthrough in particle physics. 
It is of the greatest priority to measure the properties of this new
particle with highest precision. Only high accuracy measurements may
reveal the true nature of electroweak symmetry breaking and the
fundamental structure of matter. The following questions can be
addressed at the ILC:
\begin{itemize}
\item What are the couplings of the newly discovered particle to the
  known SM particles? Are the couplings to each particle proportional to
  the particle's mass? 
\item What is the mass and width of the newly discovered particle? 
What are the spin and CP quantum numbers?
\item What is the self-coupling of the newly discovered particle?
Is the measurement consistent with the predictions from a Higgs potential?
\item Is the newly discovered particle a fundamental scalar as in the
  SM, or is only one out of several (similar) particles, potentially
  from scalar doublet, triplets, \ldots?
  Is the newly discovered particle composite?
\item Does the newly discovered particle mix with new scalars of exotic
  origin, for instance with a radion of extra-dimensional models? 
\end{itemize}
Any of these measurements may yield deviations from the SM predictions.

The Higgs boson mass plays a crucial role. Within the SM it was the last
unknown parameter and can be tested against indirect determinations, see
the next subsection. Within BSM models in which the mass can be
calculated, the measured value can be compared to this prediction, where
a precise knowledge of the top-quark mass is crucial, see \refse{sec:mt}.


\subsection{Global electroweak fits}
\label{sec:ewfit}

The precise determination of the top quark mass, together with improved
measurements of the $W$~boson mass, $\MW$, and the effective weak
leptonic mixing angle, $\sweff$, can probe the quantum corrections of
the SM and any other BSM model. The most important current and future
uncertainties are summarized in \reftas{tab:prec-exptheo},
(\ref{tab:prec-para}).

\begin{table}[ht!]
\renewcommand{\arraystretch}{1.3}
\centering
\begin{tabular}{|c||c|c||c|c||c|c|}
\hline
obs $\backslash$ prec & today & ILC/GigaZ & SM theo & SM fut & 
MSSM theo & MSSM fut \\ 
\hline\hline
$\MW$ [MeV] & 15 & $5-6$ & 4 & 1 & $5-10$ & $3-5$ \\
\hline
$\sweff$ [$10^{-5}$] & 16 & 1.3 & 4.7 & 1.5 & $5-7$ & $2.5 - 3.5$ \\
\hline
$\mt$ [GeV] & 0.9 & 0.1 & & & & \\
\hline\hline
\end{tabular}
\caption{Precision of $\MW$, $\sweff$ and $\mt$: todays experimental
  precision, future precision from ILC/GigaZ measurements, intrinsic
  uncertainties from unknown higher-order corrections in the SM today
  and in the future, in the MSSM today and in the future.}
\label{tab:prec-exptheo}
\renewcommand{\arraystretch}{1}
\end{table}

\begin{table}[h!]
\renewcommand{\arraystretch}{1.3}
\centering
\begin{tabular}{|c||c|c|c||c|c|}
\hline
obs $\backslash$ prec & $\de\mt$ & $\de\Dalhad$ & $\de\MZ$ & 
$\de\mt$,fut & $\de\Dalhad$,fut \\
\hline\hline
$\MW$ [MeV] & 5.5 & 2 & 2.5 & 1 & 1 \\
\hline
$\sweff$ [$10^{-5}$] & 7 & 3.6 & 1.4 & 0.4 & 1.8 \\
\hline\hline
\end{tabular}
\caption{Parametric uncertainties of $\MW$ and $\sweff$: todays
  experimental results using $\de\mt^{\rm exp} = 0.9 \gev$,
  $\de(\Dalhad) = 10^{-4}$ and $\de\MZ = 2.1 \mev$; future expectations
  use $\de\mt^{\rm exp,ILC} = 0.1 \gev$, 
  $\de(\Dalhad)^{\rm fut} = 5 \times 10^{-5}$.} 
\label{tab:prec-para}
\renewcommand{\arraystretch}{1}
\end{table}

\begin{figure}[htb!]
\begin{center}
\includegraphics[width=0.45\textwidth]{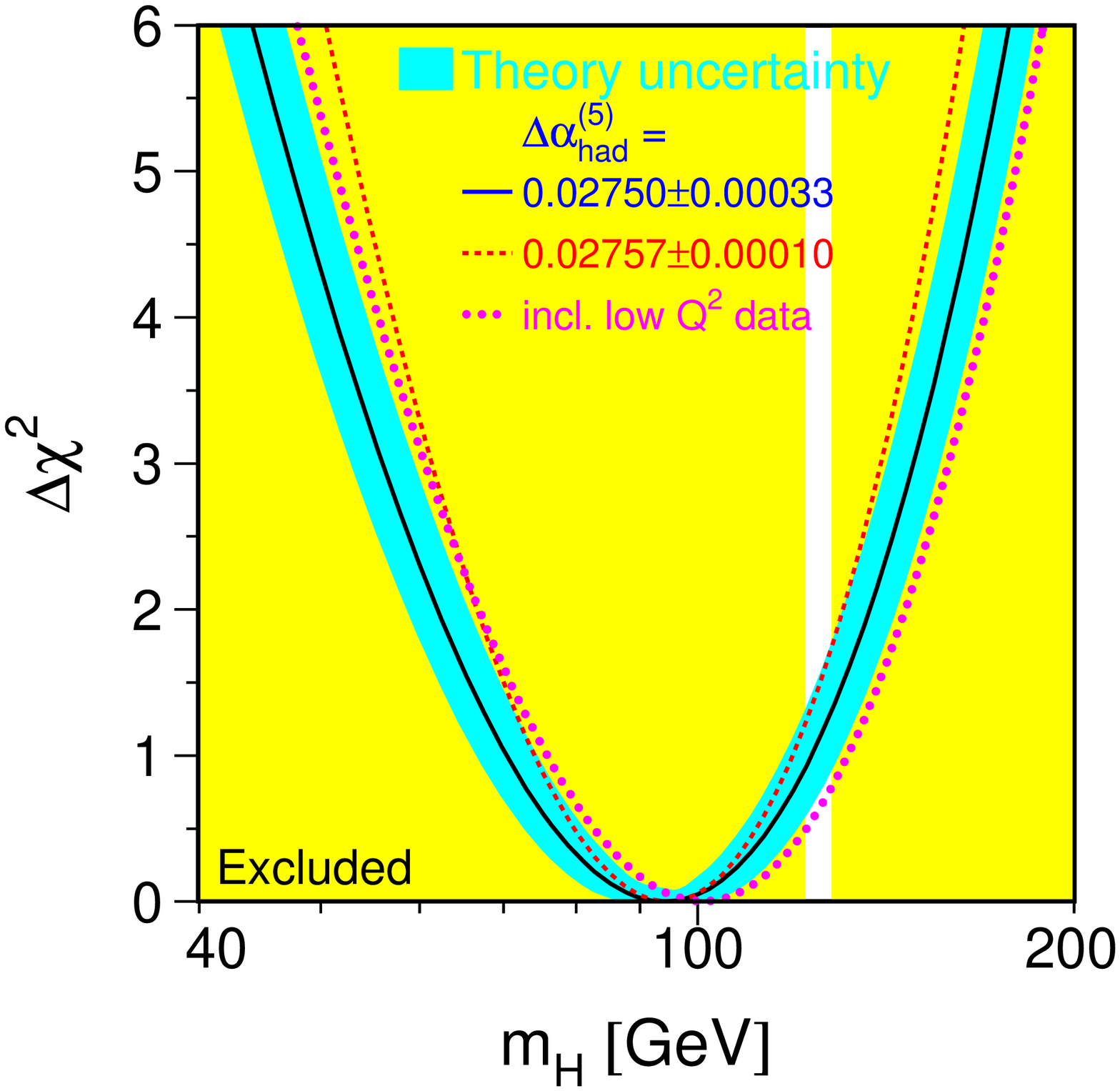}
\includegraphics[width=0.45\textwidth]{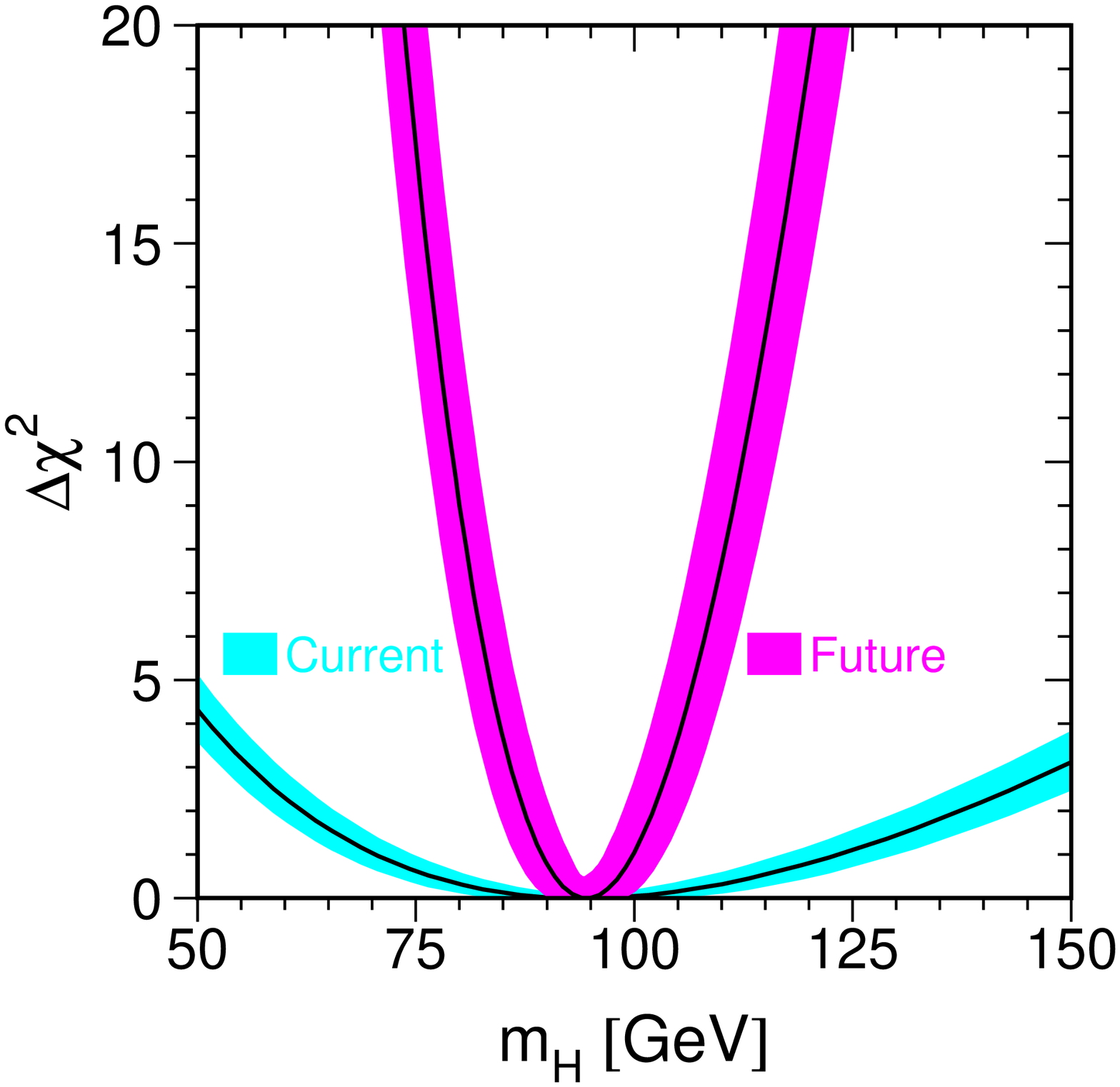}
\end{center}
\vspace{-2em}
\caption{Indirect determination of $\MH$ in the SM with current
  precision~\cite{Schael:2013ita} (left)
  and future ILC/GigaZ precision~\cite{blueband-ilcgigaz} (right plot).
}
\label{fig:blueband} 
\end{figure}

Within the SM it is possible predict the mass of the Higgs boson from
its contribution to the prediction of EWPO, see \citere{Schael:2013ita}
and references therein. With the current uncertainties%
\footnote{$\de\MW^{\rm SM,theo,fut} = 2 \mev$ has been assumed, having a minor
impact on the results.}%
~this leads
to~\cite{Schael:2013ita}
\begin{align}
\MHSM\mbox{}^{\rm \,ind} &= 94^{+29}_{-24} \gev~,
\label{MHind}
\end{align}
as it is shown in the left plot of \reffi{fig:blueband}. The left yellow
(shaded) area is excluded by LEP SM Higgs
searches~\cite{Barate:2003sz}. The right yellow (shaded) area is
excluded by LHC SM Higgs searches. 

Going to the ILC/GigaZ accuracy the
indirect determination can reach a precision of
\begin{align}
\de\MHSM\mbox{}^{\rm \,ind,ILC} &\approx \pm 10 \gev~,
\label{MHindfut}
\end{align}
as it is shown in the right plot of
\reffi{fig:blueband}~\cite{blueband-ilcgigaz}. 
Any deviation of the indirectly determined mass from the directly
measured value will indicate the presence of new physics scales beyond
the SM. Similarly, the the quantum effects in the MSSM can be tested to a very
high precision~\cite{Heinemeyer:2004gx}.


\providecommand{\Cdgz}{\ensuremath{\Delta g^\mathrm{Z}_1}}
\providecommand{\Cdgg}{\ensuremath{\Delta g^\mathrm{\gamma}_1}}
\providecommand{\Cdkz}{\ensuremath{\Delta \kappa_\mathrm{Z}}}
\providecommand{\Cdkg}{\ensuremath{\Delta \kappa_{\gamma}}}
\providecommand{\Ckg}{\ensuremath{\kappa_{\gamma}}}
\providecommand{\Ckz}{\ensuremath{\kappa_{\mathrm{Z}}}}
\providecommand{\Clg}{\ensuremath{\lambda_{\gamma}}}
\providecommand{\Clz}{\ensuremath{\lambda_{\mathrm{Z}}}}
\providecommand{\Cgv}[1]{\ensuremath{g^V_{#1}}}
\providecommand{\Cgz}[1]{\ensuremath{g^Z_{#1}}}
\providecommand{\Cgg}[1]{\ensuremath{g^{\gamma}_{#1}}}
\providecommand{\Ckzt}{\ensuremath{\tilde{\kappa}_\mathrm{Z}}}
\providecommand{\Clzt}{\ensuremath{\tilde{\lambda}_\mathrm{Z}}}
\providecommand{\Ckgt}{\ensuremath{\tilde{\kappa}_{\gamma}}}
\providecommand{\Clgt}{\ensuremath{\tilde{\lambda}_{\gamma}}}

\section{Gauge boson couplings}
\label{sec:gaugecoupl}

Another possibility to search for new physics in the electroweak sector
is the precision investigation of the couplings of the SM gauge bosons.
At the ILC at tree-level, the incoming leptons
interact via an exchange of an electroweak gauge boson. This allows
for precise studies of trilinear gauge couplings in $e^{+} e^{-} \to
W^{+} W^{-}$ as well as quartic gauge couplings occurring in a variety
of final states like $e^{+} e^{-} \to V V$ where $VV$ can be
$\gamma\gamma$ or $ZZ$, or $e^{+} e^{-} \to V V V $ with $V V V$ being
$W W Z$ or $W W \gamma$.  

One advantage of the ILC over hadron collider measurements is 
the absence of parton distribution functions such that 
the center-of-mass energy of the scattering process is 
exactly known. Together with the tunable beam energy this allows to
measure precisely the resonances.
A second advantage is the clean environment of the ILC and the
untriggered data taking. 

The trilinear electroweak gauge couplings can parametrize 
the Lagrangian~\cite{Hagiwara:1986vm}
\begin{align}
{\cal L}_{\rm TGC} &= 
         ig_{WWV}\left(g_1^V(W_{\mu\nu}^+W^{-\mu}-W^{+\mu}W_{\mu\nu}^-)V^\nu 
+\kappa^V W_\mu^+W_\nu^-V^{\mu\nu}
+\frac{\lambda^V}{\MW^2}W_\mu^{\nu+}W_\nu^{-\rho}V_\rho^{\mu}
\right. \non\\
&\; \left.
+ig_4^VW_\mu^+W^-_\nu(\partial^\mu V^\nu+\partial^\nu V^\mu)
-ig_5^V\epsilon^{\mu\nu\rho\sigma}(W_\mu^+\partial_\rho W^-_\nu-\partial_\rho W_\mu^+W^-_\nu)V_\sigma
\right.\non\\
&\; \left.
+\tilde{\kappa}^V W_\mu^+W_\nu^-\tilde{V}^{\mu\nu}
+\frac{\tilde{\lambda}^V}{\MW^2}W_\mu^{\nu+}W_\nu^{-\rho}\tilde{V}_\rho^{\mu}
\right)\;,
\label{eq:L3}
\end{align}
with
 $V=\gamma,Z$;  $W_{\mu\nu}^\pm = \partial_\mu W_\nu^\pm - \partial_\nu W_\mu^\pm$, 
 $V_{\mu\nu} = \partial_\mu V_\nu - \partial_\nu V_\mu$ and 
 $\tilde{V}_{\mu,\nu}=\epsilon_{\mu \nu \rho \sigma} V_{\rho \sigma}/2$.
(Similarly a Lagrangian for quartic gauge couplings can be
defined~\cite{Beyer:2006hx}.)
For the SM the couplings in \refeq{eq:L3} are given by
\begin{align}
 g_1^{\gamma,Z} =\kappa^{\gamma, Z} =1, \quad 
g_{4,5}^{\gamma,Z} =\tilde{\kappa}^{\gamma, Z} =1, 
 \quad \lambda^{\gamma,Z}=\tilde{\lambda}^{\gamma,Z}=0~.
\end{align}

The couplings among the electroweak gauge bosons are directly given by
the structure of the gauge group, see the previous section. 
This structure can thus directly be
determined by a measurement of the gauge boson interactions.  
Particularly sensitive is the process $e^+e^- \to W^+W^-$, since any
``naive'' change in the gauge couplings would lead to a violation of
unitarity, and small changes lead to relatively large variations.
Electroweak precision observables together with the LEP data yield the
strongest constraints on anomalous
couplings \cite{Burgess:1993vc,Aihara:1995iq}. For the triple gauge
couplings the bounds are \cite{Aihara:1995iq} 
\begin{eqnarray}
 \Delta g_1^Z &= -0.033 \pm 0.031,\non \\
 \Delta \kappa_{\gamma} &= 0.056 \pm 0.056,\non \\
 \Delta \kappa_Z &= -0.0019 \pm 0.044,\\
 \lambda_{\gamma} &= -0.036 \pm 0.034,\non \\
 \lambda_Z &= 0.049 \pm 0.045 .\non
\end{eqnarray}

Turning to the ILC, the different types of couplings can be disentangled
experimentally by analyzing the production angle distribution of the
$W$~boson and the structure of the $W$~polarization, which can be
obtained from the distributions of the decay angles.  
Anomalous couplings for $WW\gamma$ and $WWZ$ 
result in similar final state distributions. However, using beam
polarization, they can be disentangled, where a large beam polarization,
in particular for the left-handed $e^-$ is required. Also positron
polarization is required for an optimal
resolution~\cite{MoortgatPick:2005cw}. 
A fast detector simulation analysis was performed for 
$\sqrt{s} = 500 \gev$ and $800 \gev$~\cite{Menges}.
The results for single parameter fits are shown in 
\refta{tab:tgc}. Correlations in the multi-parameter fits were taken
into account where possible. For $\sqrt{s} = 800 \gev$ 
they are relatively small, not increasing the uncertainties by more than
$\sim 20\%$. At $\sqrt{s} = 500 \gev$ the effect is larger, and
uncertainties can increase by up to a factor of two, see
also \citere{AguilarSaavedra:2001rg}. 

\begin{figure}[htb!]
  \centering
  \includegraphics[width=0.4\linewidth,bb=33 17 492 468]{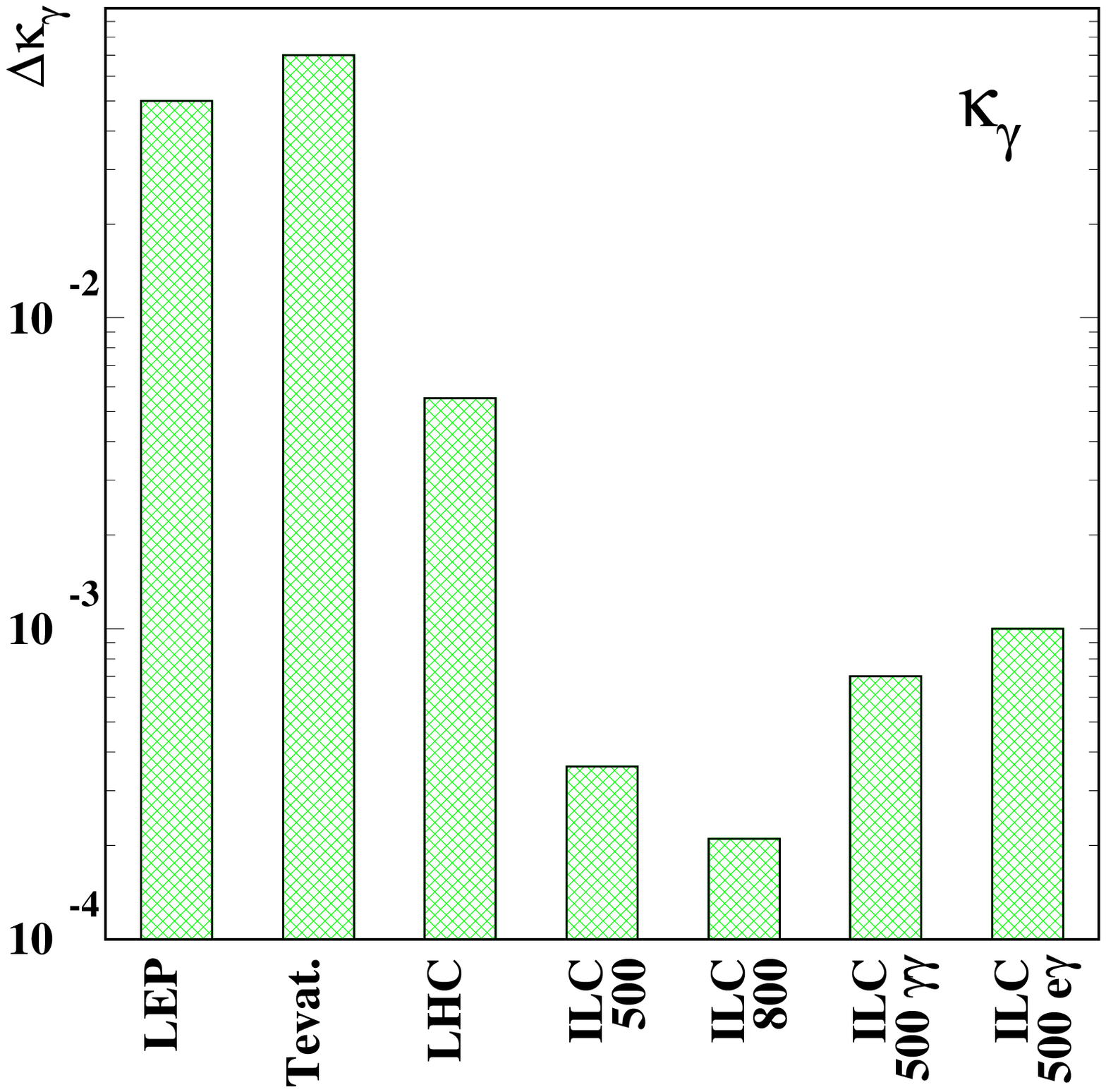}
  \hspace*{5mm}
  \includegraphics[width=0.4\linewidth,bb=33 17 492 468]{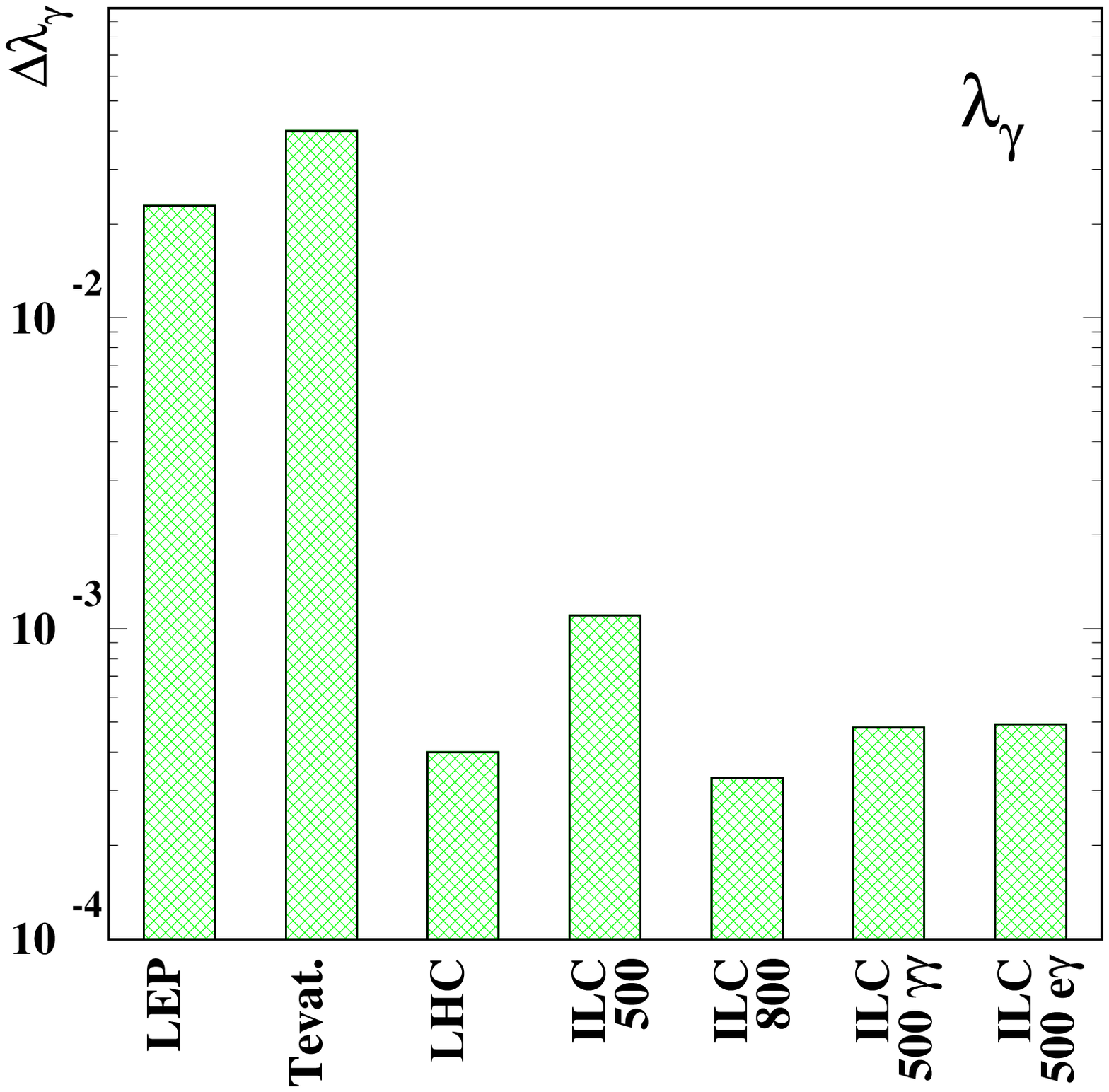}
\vspace*{-5mm}
  \caption{
    Comparison of $\Delta \kappa_\gamma$ and $\Delta
      \lambda_\gamma$ at different machines. For LHC and ILC three
      years of running are assumed (LHC: $300\ifb$, ILC
      $\sqrt{s}=500\gev$: $500 \ifb$, ILC $\sqrt{s}=800\gev$: $1000
      \ifb$). If available the results from multi-parameter fits have
      been used. Taken from \citere{ilc:rdr}.}
 \label{fig:tgc}
\end{figure}

Figure \ref{fig:tgc} compares the expected precision for the $\kappa_\gamma$
and $\lambda_\gamma$ measurements at various colliders \cite{ilc:rdr}. The
advantage of ILC is clearly seen for the $\kappa$ couplings. It is an example
of new physics effects with lower mass dimension operators, for which precise
measurements at low energies can be more effective than less precise
measurements at higher energies.

\begin{table}
\centering
\renewcommand{\arraystretch}{1.2}
\begin{tabular}[c]{|c|c|c|}
\hline
coupling & \multicolumn{2}{|c|}{error $\times 10^{-4}$} \\
\cline{2-3}
         & $\sqrt{s}=500\gev$ & $\sqrt{s}=800\gev$ \\
\hline
\hline
  \Cdgz  &$ 15.5 \phantom{0} $&$ 12.6 \phantom{0} $\\
  \Cdkg  &$  3.3 $&$  1.9 $\\
  \Clg   &$  5.9 $&$  3.3 $\\
  \Cdkz  &$  3.2 $&$  1.9 $\\
  \Clz   &$  6.7 $&$  3.0 $\\
\hline
\hline         
  \Cgz{5}&$ 16.5 \phantom{0} $&$ 14.4 \phantom{0} $\\
  \Cgz{4}&$ 45.9 \phantom{0} $&$ 18.3 \phantom{0} $\\
  \Ckzt  &$ 39.0 \phantom{0} $&$ 14.3 \phantom{0} $\\
  \Clzt  &$  7.5 $&$  3.0 $\\
  \hline
\end{tabular}
\caption{
  Results of the single parameter fits ($1 \sigma$) to the different 
    triple gauge couplings at the ILC for $\sqrt{s}=500 \gev$ with ${\cal L}=
    500\ifb$ and $\sqrt{s}=800 \gev$ with ${\cal L}=1000\ifb$;
    ${\cal P}_{e^-} = 80\%$ and ${\cal P}_{e^+} = 60\%$ has been used.}
\label{tab:tgc} 
\end{table}

\newcommand{\comm}[1]{{\color{blue} [#1]}}

\section{Extra gauge bosons}
\label{sec:extragaugebosons}

 The two-fermion processes $e^+ e^- \rightarrow f \bar f$, where $f$ is a quark or charged lepton, are especially powerful
 probes of TeV scale $Z'$ s,   extra dimensions, and other new physics
 leading to contact interactions such as quark or lepton compositeness.
  
 \subsection{\boldmath{$Z'$} Gauge Bosons}
 Additional $U(1)'$ gauge symmetries occur in many extensions of the standard model (SM), often 
 with the $Z'$ mass at the TeV scale\footnote{For reviews, see, e.g.,~\cite{Cvetic:1995zs,Hewett:1988xc,Leike:1998wr,Langacker:2008yv,Nath:2010zj}.
 For previous studies in $e^+e^-$ colliders, see~\cite{DelAguila:1993rw,DelAguila:1995fa,Accomando:1997wt,Weiglein:2004hn,Godfrey:2005pm,Osland:2009dp,Linssen:2012hp,TDR_Physics,Battaglia:2012ez}.}. 
 For example, grand unified theories
 and  string constructions often involve large underlying gauge groups, which can easily leave behind
 (remnant) $U(1)'$ factors in addition to the SM group when broken. Some string constructions,
 such as the type IIa intersecting-brane theories, are based on $U(n) = SU(n) \times U(1)$
 factors. Although the $U(1)$ factors are typically anomalous, linear combinations (in addition to weak
 hypercharge) may be non-anomalous and survive to low energies. In supersymmetric models
 with an additional $U(1)'$ both the $SU(2) \times U(1)$ and $U(1)'$ breaking scales
 are typically set by the supersymmetry-breaking soft parameters (unless there are flat directions),
 and the $U(1)'$ can provide an elegant solution to the $\mu$ problem similar to the NMSSM.
 
 $U(1)'$ s also occur in extended electroweak models, such as left-right symmetry, and in many alternative models of electroweak symmetry breaking (EWSB),
 such as various dynamical symmetry breaking  or  Little Higgs models. In some cases these feature enhanced couplings to the third generation.
 Although many of the alternative EWSB models  are disfavored by the discovery  of the Higgs-like boson at
 $\sim$125 GeV, there still remains the possibility that the minimal Higgs model is only an approximation to
 an underlying alternative mechanism.  Models in which the  photon and $Z$ propagate in extra dimensions   involve Kaluza-Klein excitations, which could resemble a $Z'$. 
 For a flat dimension of radius $R$, for example, the Kaluza-Klein mass scale is $M\sim R^{-1}\sim 2 {\ \rm TeV }\times
(10^{-17} {\rm cm }/R)$.
 Extra (family non-universal) $Z'$ s have also been invoked in connection with
 possible experimental anomalies, such as the still-unexplained forward-backward asymmetry in $t \bar t$ production reported by CDF and D0.
 
$Z'$ s have also been motivated by  other considerations, such as light weakly-coupled $Z'$ s which could communicate
 with an otherwise dark sector. However, we concentrate on TeV-scale $Z'$ s with electroweak-strength couplings, for which the ILC
 has a significant reach. We emphasize that the observation of such a $Z'$ would have major consequences: most models are
 accompanied by extended Higgs and neutralino sectors and  new exotic fermions (for anomaly cancellation), and may have implications for neutrino mass,
 electroweak baryogenesis, tree-level FCNC, and the mediation of supersymmetry breaking. An on-shell $Z'$ could serve as a sparticle/exotics factory.
 
 The effects of $Z'$ s have been extensively searched for in precision electroweak physics, which limits any $Z-Z'$ mixing to a few parts in 
 a thousand and places lower limits on $M_{Z'}$ ranging from a few hundred GeV to $\sim$1 TeV for typical benchmark models~\cite{Erler:2009jh,Diener:2011jt}.
 The mass limits were superseded by direct searches at the Tevatron and then the LHC for resonant $Z'$ production with decays into dileptons or into
 $t \bar t$  or $b \bar b$ pairs. The current ATLAS and CMS lower limits from dileptons are in the 2-2.5 TeV range for benchmark models, with a future
 discovery reach of $\sim 4-5$ TeV for $\sqrt{s}=14$ TeV and $\int \mathcal{L}=100$ fm$^{-1}$. For lower masses (up to around 2.5 TeV) there would be significant possibilities
 for discriminating between $Z'$ models at the LHC utilizing forward-backward asymmetries, rapidity distributions, lineshape variables, other decay modes, $\tau$ polarization,
 associated production, and rare decays. However, for the larger masses favored by the present LHC constraints the diagnostic possibilites are more limited\footnote{The LHC possibilities are extensively discussed in~\cite{Langacker:2008yv,Nath:2010zj,DelAguila:1993rw,DelAguila:1995fa,Weiglein:2004hn,Langacker:1984dc,delAguila:1993ym,Dittmar:2003ir,Kang:2004bz,Petriello:2008zr,Godfrey:2008vf,Osland:2009tn,Li:2009xh,Diener:2009vq,delAguila:2010mx,Chang:2011be,Erler:2011ud,Chiang:2011kq}.}.
 
The ILC has a significant reach for observing the effects of a $Z'$ and discriminating
between models due to the interference of a heavy virtual\footnote{Existing limits already exclude $Z'$ s with electroweak couplings to charged leptons with mass light enough for resonant production at the ILC.} $Z'$ with $s$-channel $\gamma$ and $Z$ exchange
in the processes $e^+ e^- \rightarrow f \bar f$, with $f=e, \mu, \tau, c,$ and $b$ ($t$-channel exchange must also be considered for $f=e$). For unpolarized beams the basic observables are the total cross sections, forward backward asymmetries, and $\tau$ polarization. Additional probes, including the polarization asymmetry and the forward-backward polarization asymmetry,
become available for polarized $e^\pm$. The ability to polarize both beams leads to a larger effective polarization as well as
providing an additional handle for new-physics identification and systematics effects.

\begin{figure}[htbp]
\mbox{}\vspace{-10em}
\begin{center}
\includegraphics*[scale=0.8]{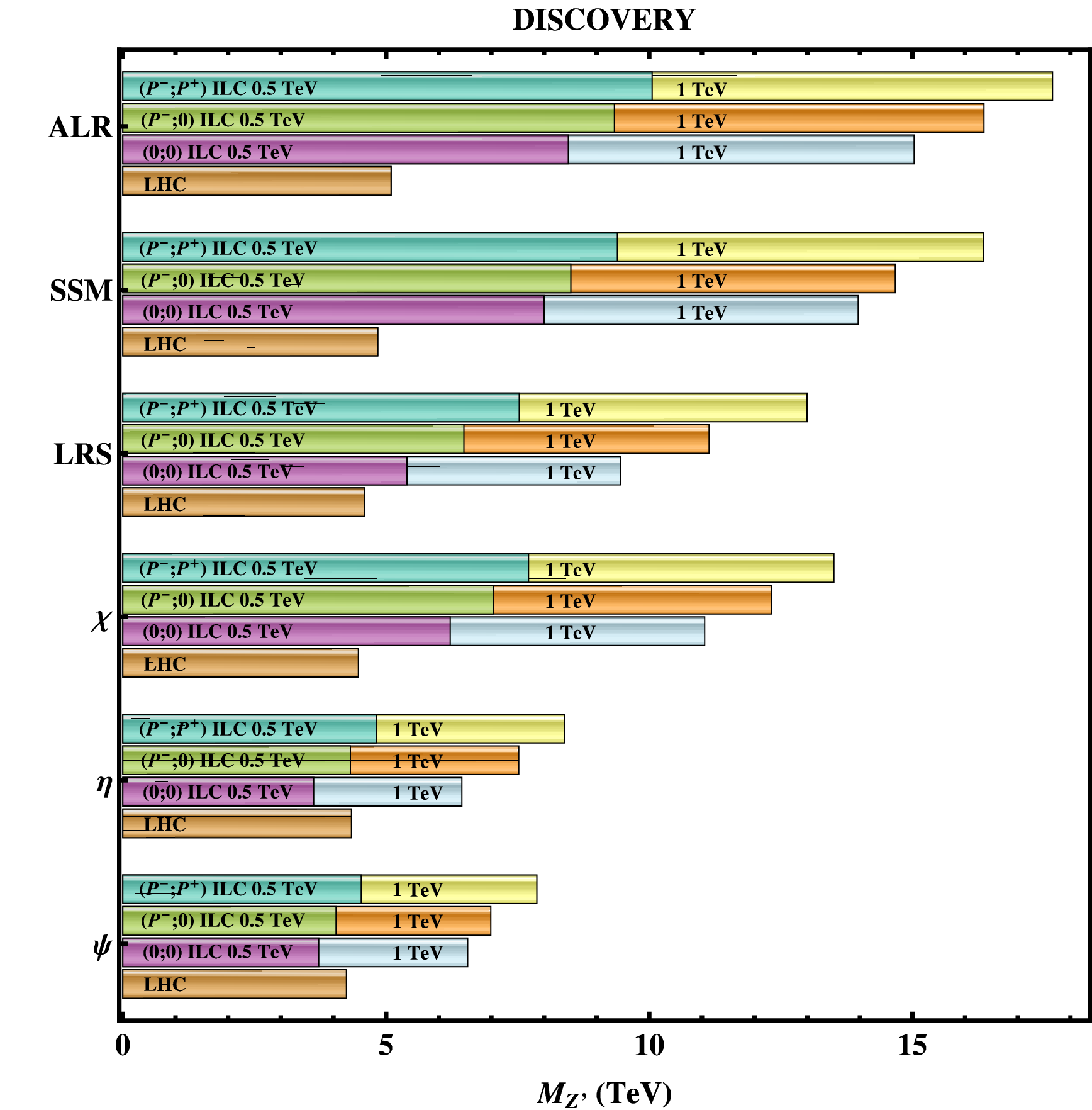}
\caption{95\% C.L. discovery reach of the ILC  for various benchmark $Z'$ models at $\sqrt{s}=$0.5 TeV  (1 TeV),
integrated luminosity 500 fb$^{-1}$ (1000 fb$^{-1}$), and various polarization scenarios, from~\cite{Osland:2009dp}. The nonzero
polarizations are $|P^-|=0.8$ and $|P^+|=0.6$. Also shown is the LHC 5$\sigma$
reach at $\sqrt{s}=14$ TeV and $\int \mathcal{L}=100$ fb$^{-1}$.}
\label{zpreach}
\end{center}
\vspace{3em}
\end{figure}

Especially detailed studies of the ILC discovery potential and model discrimination have been carried out in~\cite{Godfrey:2005pm} and~\cite{Osland:2009dp}. The 95\% C.L. discovery reach for the ILC for various benchmark models is shown
in Figure~\ref{zpreach}, taken from~\cite{Osland:2009dp}. It is seen that the reach for $\sqrt{s}=$0.5 TeV 
 and $\int \mathcal{L}=$ 500 fb$^{-1}$ is in the range  4-10 TeV for the models considered, increasing with  beam polarization.
 For 1 TeV and 1000 fb$^{-1}$ the range increases to around 6-15 TeV, well above the reach of the LHC.
 
\begin{figure}[htbp]
\mbox{}\vspace{-10em}
\begin{center}
\includegraphics*[scale=0.8]{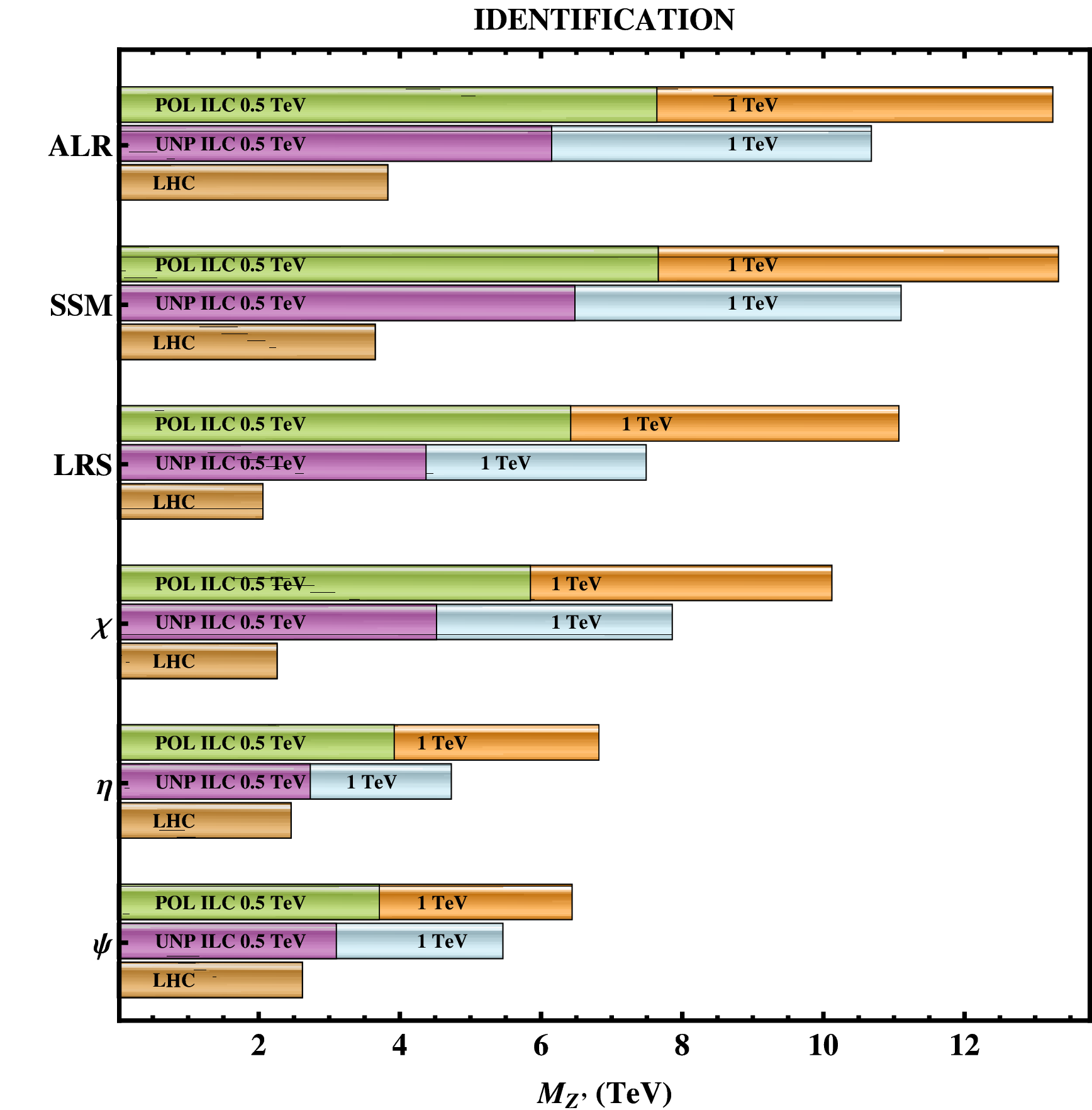}
\caption{Reach for discriminating between the models shown at 95\% C.L., from~\cite{Osland:2009dp}. The energies and integrated luminosities
are as in Figure~\ref{zpreach}, while UNP and POL refer respectively to no polarization, and to $|P^-|=0.8$, $|P^+|=0.6$.}
\label{zpid}
\end{center}
\vspace{3em}
\end{figure}

 The ILC also would allow excellent discrimination between $Z'$ models. Of course, the cleanest identification would be for a relatively low mass $Z'$ that had already been observed (and its mass determined) at the LHC. However, 
the diagnostic reach of the ILC extends much higher, 
almost up to the discovery reach, and far beyond the possibilities for the LHC. This is illustrated in Figure~\ref{zpid}, taken from~\cite{Osland:2009dp}.
  
\clearpage
\subsection{\boldmath{$W'$} Gauge Bosons}
\label{sec:wprime}
 
 Heavy singly charged gauge bosons $W'$ occur, for example, in
$SU(2)_L \times SU(2)_R \times U(1)$ models (including left-right symmetric models),
in many extended models of electroweak symmetry breaking (e.g., involving $SU(2)_1 \times SU(2)_2$
breaking to a diagonal subgroup), and as Kaluza-Klein excitations in models in which the
$W$ propagates in extra dimensions\footnote{For recent general discussions, see, e.g.,~\cite{Gopalakrishna:2010xm,Frank:2010cj,Schmaltz:2010xr}, which contain references to earlier papers.}.

$W'$ couplings to the SM and other fermions are characterized by chirality (i.e., whether the couplings are to $V+A$,
as in $SU(2)_L \times SU(2)_R \times U(1)$, or to $V-A$, as in the diagonal embeddings of $SU(2)$ or for Kaluza-Klein excitations); the analog of the CKM matrix for the quark currents; possible $W-W'$ mixing; and the $W'$ gauge couplings. For $V+A$ couplings one must also specify whether the right-handed neutrinos $\nu_R$ are light (as for Dirac neutrinos or possibly for eV-scale ``sterile'' neutrinos) or heavy (as in an ordinary seesaw model).
In the latter case, the $\nu_R$ could be TeV scale or lighter, or could be too heavy to produce at the LHC or other facilities.

The existing constraints are difficult to summarize because of the many possibilities mentioned above.
There are significant limits on mass and mixing from $\beta$ and $\mu$ decay and weak universality
(from $W'$ exchange and $W-W'$ mixing), from neutral meson mixing (from box diagrams),
from the $W$ mass and other precision electroweak, and from correlated $W'-Z'$ constraints in specific models.
Depending on the couplings, these set lower limits on $M_{W'}$ ranging from several hundred GeV to around 2.5 TeV~\cite{delAguila:2010mx,Gopalakrishna:2010xm,Frank:2010cj,Schmaltz:2010xr,Hsieh:2010zr,Frank:2010qv}. 

At the Tevatron and LHC~\cite{Gopalakrishna:2010xm,Frank:2010cj,Schmaltz:2010xr,Cao:2012ng,Du:2012vh,Duffty:2012rf} one can search for $W' \rightarrow \ell \nu$ for $V-A$ (or $V+A$ with light $\nu_R$),
or for $W' \rightarrow \ell \nu_R, \nu_R \rightarrow \ell j j$ for a  heavy $\nu_R$ that is lighter than the $W'$.
In all cases one can also utilize the nonleptonic decays such as $W'\rightarrow t \bar b$.
The current CMS and ATLAS lower  limits on $M_{W'}$
are around 3.35 TeV for $ \ell \nu$~\cite{CMS12060,Aad:2012dm}, 2.5 TeV for  kinematically allowed leptonic decays with a heavy neutrino~\cite{ATLAS:2012ak,CMS12017}, and 1.8 TeV for
non-leptonic decays~\cite{Chatrchyan:2012gqa,Atlas2013050}. Most of these limits are based on only a fraction of the
existing data.
Higher energy and luminosity will presumably allow considerably improved sensitivity into the multi-TeV range. Additional information on the chirality, etc., may be obtained by leptonic decays, top polarization~\cite{Gopalakrishna:2010xm},
associated $W' t$ production~\cite{Frank:2010cj}, etc. 

 The ILC offers only limited possibilities for discovery or diagnostics of a $W'$. In particular,
 the process $e^- e^+ \rightarrow \nu \bar\nu \gamma$ could proceed via $t$-channel $W'$ exchange
 (with the photon radiated from a charged particle). The sensitivity extends to around 6 TeV for $V-A$ and 1.9 TeV for $V+A$ with a light Dirac neutrino, assuming $\sqrt{s}=1$ TeV, $\int \mathcal{L}=$ 500 fb$^{-1}$,  high beam polarizations, and favorable assumptions concerning the coupling strength~\cite{Godfrey:2000hc}.
 Although not competitive with the LHC for discovery, measurements of the cross section and left-right asymmetry in 
 $ \nu \bar\nu \gamma$ could help constrain the chirality.


\section{Conclusions}
\label{sec:concl}

We have reviewed the experimental conditions and prospects for the
measurement of precision observables at the ILC. 

The observables comprise the $W$~boson mass, $Z$~pole observables and in
particular the effective weak leptonic mixing angle, $\sweff$, the
top-quark mass (and other top related observables), the strong coupling
constant, $\als$, and the Higgs boson mass. These measurement of these
observables will allow a test of the SM (or any other model) at the
quantum level. In particular, the Higgs boson mass can be determined
{\em indirectly}, which can be compared to the {\em directly} measured
value. 

In the case that the LHC will discover no additional new physics,
i.e.\ only the full particle content of the SM would be discovered, the
precision measurements at the ILC of the Higgs boson properties (in
particular the couplings to other SM particles and to itself), together
with the precision determinations of $\mt$, $\MW$, $\sweff$ etc.\ would
constitute a clear way to search for new scales beyond the SM. Only the
high anticipated precision would offer the opportunity to find traces of
high scales beyond the direct kinematical reach of the LHC or the ILC.

We have also reviewed the prospects for the measurements of triple and
quartic (SM) gauge boson couplings as well as the potential measurement
of additional heavy gauge bosons. While currently no deviations in the
couplings from the SM values can be observed, these measurements have
the potential to find traces of physics beyond the SM. Scales even
beyond $10 \tev$ can be probed in models with additional gauge bosons,
far beyond the direct kinematical reach of the LHC or the ILC.


\subsection*{Acknowledgements}

We acknowledge useful discussions with 
M.~Gr\"unewald, 
A~H\"ocker 
and
R.~Kogler. 
The work of S.H.\ was supported in part by CICYT 
(grant FPA 2010--22163-C02-01) and by the Spanish MICINN's 
Consolider-Ingenio 2010 Program under grant MultiDark CSD2009-00064.


\newpage
\pagebreak


\begin{thebibliography}{99} 


\bibitem{TDR} 
Technical Design Report, The ILC Baseline Design, \\ 
{\tt https://forge.linearcollider.org/dist/20121210-CA-TDR2.pdf}~(2013).

\bibitem{TDR_Physics} H.~Baer {\it et al.},
``The International Linear Collider Technical Design Report - Volume 2:
  Physics``, 
  arXiv:1306.6352 [hep-ex].

\bibitem{TDR_DBDs}
ILC Detector Baseline Document, 
{\tt http://ific.uv.es/~fuster/DBD-Chapters/}~(2013).

\bibitem{ILC_Parameters_Document}
Parameters for the Linear Collider, \newline 
{\tt http://www.fnal.gov/directorate/icfa/recent\_lc\_activities\_files/\\
para-Nov20-final.pdf} 
(2003, updated 2006).

\bibitem{MoortgatPick:2005cw}
  G.~Moortgat-Pick, T.~Abe, G.~Alexander, B.~Ananthanarayan, A.~A.~Babich,
V.~Bharadwaj, D.~Barber and A.~Bartl {\it et al.},
  Phys.\ Rept.\  {\bf 460} (2008) 131
  [hep-ph/0507011].

\bibitem{List} M. Beckmann, B. List, J. List, NIM paper.

\bibitem{Moenig-and-Sailer-Diplomarbeit}
K.~Moenig, LC-PHSM-2000-060;\\
A.~Sailer, Diploma Thesis. 

\bibitem{Group:2012gb}
  Tevatron Electroweak Working Group [CDF and D0 Collaborations],
  arXiv:1204.0042 [hep-ex].

\bibitem{Alcaraz:2006mx}
  J.~Alcaraz {\it et al.}  [ALEPH and DELPHI and L3 and OPAL and LEP Electroweak Working Group Collaborations],
  hep-ex/0612034.

\bibitem{Wilson_Sitges} G.W.~Wilson, LC-PHSM-2001-009, see:
  {\tt http://www-flc.desy.de/lcnotes/}~.

\bibitem{Denner:2005es}
  A.~Denner, S.~Dittmaier, M.~Roth and L.~H.~Wieders,
  Phys.\ Lett.\ B {\bf 612} (2005) 223
   [Erratum-ibid.\ B {\bf 704} (2011) 667]
  [hep-ph/0502063].

\bibitem{Denner:2005fg}
  A.~Denner, S.~Dittmaier, M.~Roth and L.~H.~Wieders,
  Nucl.\ Phys.\ B {\bf 724} (2005) 247
   [Erratum-ibid.\ B {\bf 854} (2012) 504]
  [hep-ph/0505042].

\bibitem{Baur:2001yp}
  U.~Baur, R.~Clare, J.~Erler, S.~Heinemeyer, D.~Wackeroth, G.~Weiglein
  and D.~R.~Wood, 
  eConf C {\bf 010630} (2001) P122
  [hep-ph/0111314].

\bibitem{Heinemeyer:2004gx}
  S.~Heinemeyer, W.~Hollik and G.~Weiglein,
  Phys.\ Rept.\  {\bf 425} (2006) 265
  [hep-ph/0412214].

\bibitem{deMWSMtheo} M.~Awramik, M.~Czakon, A.~Freitas and G.~Weiglein,
                     Phys.\ Rev.\ D {\bf 69} (2004) 053006
                     [hep-ph/0311148].

\bibitem{Heinemeyer:2006px}
  S.~Heinemeyer, W.~Hollik, D.~Stockinger, A.~M.~Weber and G.~Weiglein,
  JHEP {\bf 0608} (2006) 052
  [arXiv:hep-ph/0604147].

\bibitem{MWlisa} S.~Heinemeyer, G.~Weiglein and L.~Zeune,
                 DESY 13--015,
                 {\em in preparation}.

\bibitem{gigaz} R.~Hawkings and K.~M\"onig,
                EPJdirect C {\bf 8} (1999) 1
                [arXiv:hep-ex/9910022].

\bibitem{gigazsitges} S.~Heinemeyer, Th.~Mannel  and G.~Weiglein,
                      Proceedings, Linear Collider Workshop Sitges 1999,
                      arXiv:hep-ph/9909538.

\bibitem{lepz}
  S.~Schael {\it et al.}  [ALEPH and DELPHI and L3 and OPAL and SLD and LEP
    Electroweak Working Group and SLD Electroweak Group and SLD Heavy Flavour
    Group Collaborations], 
  Phys.\ Rept.\  {\bf 427} (2006) 257
  [hep-ex/0509008].

\bibitem{Blondel:1987wr}
  A.~Blondel,
  Phys.\ Lett.\ B {\bf 202} (1988) 145
  [Erratum-ibid.\  {\bf 208} (1988) 531].

\bibitem{sw2effSM}  M.~Awramik, M.~Czakon and A.~Freitas,
  Phys.\ Lett.\  B {\bf 642}, 563 (2006)
  [hep-ph/0605339];\\
  W.~Hollik, U.~Meier and S.~Uccirati,
  Nucl.\ Phys.\ B {\bf 765}, 154 (2007)
  [hep-ph/0610312].

\bibitem{sw2effSM2} M.~Awramik, M.~Czakon and A.~Freitas,
  JHEP {\bf 0611}, 048 (2006)
   [hep-ph/0608099].

\bibitem{Heinemeyer:2007bw}
  S.~Heinemeyer, W.~Hollik, A.~M.~Weber and G.~Weiglein,
  JHEP {\bf 0804} (2008) 039
  [arXiv:0710.2972 [hep-ph]].

  \bibitem{Barate:2003sz}
  R.~Barate {\it et al.}  [LEP Working Group for Higgs boson searches and ALEPH and DELPHI and L3 and OPAL Collaborations],
  Phys.\ Lett.\ B {\bf 565} (2003) 61
  [hep-ex/0306033].

\bibitem{gruenewaldpriv} M.~Gr\"unewald, 
                         {\em private communication}.

\bibitem{swbb}
  M.~Awramik, M.~Czakon, A.~Freitas and B.~A.~Kniehl,
  Nucl.\ Phys.\ B {\bf 813} (2009) 174
  [arXiv:0811.1364 [hep-ph]].

\bibitem{bnew}
  B.~Batell, S.~Gori and L.-T.~Wang,
  JHEP {\bf 1301} (2013) 139
  [arXiv:1209.6382 [hep-ph]];\\
  D.~Choudhury, A.~Kundu and P.~Saha,
  arXiv:1305.7199 [hep-ph];\\
  M.~Ciuchini, E.~Franco, S.~Mishima and L.~Silvestrini,
  arXiv:1306.4644 [hep-ph].

\bibitem{rb}
  A.~Freitas and Y.-C.~Huang,
  JHEP {\bf 1208} (2012) 050
  [Erratum-ibid.\  {\bf 1305} (2013) 074]
  [arXiv:1205.0299 [hep-ph]].

\bibitem{gzmt}
  J.~Fleischer, O.~V.~Tarasov and F.~Jegerlehner,
  Phys.\ Rev.\ D {\bf 51} (1995) 3820;\\
  G.~Degrassi and P.~Gambino,
  Nucl.\ Phys.\ B {\bf 567} (2000) 3
  [hep-ph/9905472].

\bibitem{Lancaster:2011wr}
  [Tevatron Electroweak Working Group and CDF and D0 Collaborations],
  arXiv:1107.5255 [hep-ex];
  arXiv:1305.3929 [hep-ex].

\bibitem{Degrassi:2002fi}
  G.~Degrassi, S.~Heinemeyer, W.~Hollik, P.~Slavich and G.~Weiglein,
  Eur.\ Phys.\ J.\ C {\bf 28} (2003) 133
  [hep-ph/0212020].

\bibitem{Erler:2000jg}
  J.~Erler, S.~Heinemeyer, W.~Hollik, G.~Weiglein and P.~M.~Zerwas,
  Phys.\ Lett.\ B {\bf 486} (2000) 125
  [hep-ph/0005024].

\bibitem{Schael:2013ita}
  S.~Schael {\it et al.}  [ALEPH and DELPHI and L3 and OPAL and LEP Electroweak Working Group Collaborations],
  arXiv:1302.3415 [hep-ex].

\bibitem{blueband-ilcgigaz} M.~Gr\"unewald, 
                            {\em priv. communication};\\
                            G.~Moortgat-Pick {\it et al.}, 
                            ``LC review'', to appear in
                            Eur.\ Phys.\ J.\ {\bf C}~.

\bibitem{Hagiwara:1986vm} 
  K.~Hagiwara, R.~D.~Peccei, D.~Zeppenfeld and K.~Hikasa,
  Nucl.\ Phys.\ B {\bf 282} (1987) 253.

\bibitem{Beyer:2006hx}
  M.~Beyer, W.~Kilian, P.~Krstonosic, K.~Monig, J.~Reuter, E.~Schmidt and
  H.~Schroder, 
  Eur.\ Phys.\ J.\ C {\bf 48} (2006) 353
  [hep-ph/0604048].

\bibitem{Burgess:1993vc}
C.~P.~Burgess, S.~Godfrey, H.~Konig, D.~London and I.~Maksymyk,
  Phys.\ Rev.\ D {\bf 49} (1994) 6115
  [arXiv:hep-ph/9312291].

\bibitem{Aihara:1995iq}
  H.~Aihara, T.~Barklow, U.~Baur, J.~Busenitz, S.~Errede, T.~A.~Fuess, T.~Han
  and D.~London {\it et al.}, 
  [arXiv:hep-ph/9503425].

\bibitem{Menges} W.~Menges, LC-PHSM-2001-022, 
                  see: {\tt www-flc.desy.de/lcnotes/}~.

\bibitem{AguilarSaavedra:2001rg}
  J.~A.~Aguilar-Saavedra {\it et al.}  [ECFA/DESY LC Physics Working Group
  Collaboration], 
  hep-ph/0106315.

\bibitem{ilc:rdr} 
  G.~Aarons {\it et al.}  [ILC Collaboration],
  arXiv:0709.1893 [hep-ph].

\bibitem{Cvetic:1995zs}
  M.~Cvetic and S.~Godfrey,
  In *Barklow, T.L. (ed.) et al.: Electroweak symmetry breaking and new physics at the TeV scale* 383-415
  [hep-ph/9504216].

\bibitem{Hewett:1988xc}
  J.~L.~Hewett and T.~G.~Rizzo,
  Phys.\ Rept.\  {\bf 183} (1989) 193.

\bibitem{Leike:1998wr}
  A.~Leike,
  Phys.\ Rept.\  {\bf 317} (1999) 143
  [hep-ph/9805494].

\bibitem{Langacker:2008yv}
  P.~Langacker,
  Rev.\ Mod.\ Phys.\  {\bf 81} (2009) 1199
  [arXiv:0801.1345 [hep-ph]].

\bibitem{Nath:2010zj}
  P.~Nath, B.~D.~Nelson, H.~Davoudiasl, B.~Dutta, D.~Feldman, Z.~Liu, T.~Han and P.~Langacker {\it et al.},
  Nucl.\ Phys.\ Proc.\ Suppl.\  {\bf 200-202} (2010) 185
  [arXiv:1001.2693 [hep-ph]].

\bibitem{DelAguila:1993rw}
  F.~Del Aguila and M.~Cvetic,
  Phys.\ Rev.\ D {\bf 50} (1994) 3158
  [hep-ph/9312329].

  \bibitem{DelAguila:1995fa}
  F.~Del Aguila, M.~Cvetic and P.~Langacker,
  Phys.\ Rev.\ D {\bf 52} (1995) 37
  [hep-ph/9501390].

\bibitem{Accomando:1997wt}
  E.~Accomando {\it et al.}  [ECFA/DESY LC Physics Working Group Collaboration],
  Phys.\ Rept.\  {\bf 299} (1998) 1
  [hep-ph/9705442].

\bibitem{Weiglein:2004hn}
  G.~Weiglein {\it et al.}  [LHC/LC Study Group Collaboration],
  Phys.\ Rept.\  {\bf 426} (2006) 47
  [hep-ph/0410364].

\bibitem{Godfrey:2005pm}
  S.~Godfrey, P.~Kalyniak and A.~Tomkins,
  hep-ph/0511335.

\bibitem{Osland:2009dp}
  P.~Osland, A.~A.~Pankov and A.~V.~Tsytrinov,
  Eur.\ Phys.\ J.\ C {\bf 67} (2010) 191
  [arXiv:0912.2806 [hep-ph]].

\bibitem{Linssen:2012hp}
  L.~Linssen, A.~Miyamoto, M.~Stanitzki and H.~Weerts,
  arXiv:1202.5940 [physics.ins-det].

\bibitem{Battaglia:2012ez}
  M.~Battaglia, F.~Coradeschi, S.~De Curtis and D.~Dominici,
  arXiv:1203.0416 [hep-ph].

\bibitem{Erler:2009jh}
  J.~Erler, P.~Langacker, S.~Munir and E.~Rojas,
  JHEP {\bf 0908} (2009) 017
  [arXiv:0906.2435 [hep-ph]].

\bibitem{Diener:2011jt}
  R.~Diener, S.~Godfrey and I.~Turan,
  Phys.\ Rev.\ D {\bf 86} (2012) 115017
  [arXiv:1111.4566 [hep-ph]].

\bibitem{Langacker:1984dc}
  P.~Langacker, R.~W.~Robinett and J.~L.~Rosner,
  Phys.\ Rev.\ D {\bf 30} (1984) 1470.

\bibitem{delAguila:1993ym}
  F.~del Aguila, M.~Cvetic and P.~Langacker,
  Phys.\ Rev.\ D {\bf 48} (1993) 969
  [hep-ph/9303299].

\bibitem{Dittmar:2003ir}
  M.~Dittmar, A.~-S.~Nicollerat and A.~Djouadi,
  Phys.\ Lett.\ B {\bf 583} (2004) 111
  [hep-ph/0307020].

\bibitem{Kang:2004bz}
  J.~Kang and P.~Langacker,
  Phys.\ Rev.\ D {\bf 71} (2005) 035014
  [hep-ph/0412190].

\bibitem{Petriello:2008zr}
  F.~Petriello and S.~Quackenbush,
  Phys.\ Rev.\ D {\bf 77} (2008) 115004
  [arXiv:0801.4389 [hep-ph]].

\bibitem{Godfrey:2008vf}
  S.~Godfrey and T.~A.~W.~Martin,
  Phys.\ Rev.\ Lett.\  {\bf 101} (2008) 151803
  [arXiv:0807.1080 [hep-ph]].

\bibitem{Osland:2009tn}
  P.~Osland, A.~A.~Pankov, A.~V.~Tsytrinov and N.~Paver,
  Phys.\ Rev.\ D {\bf 79} (2009) 115021
  [arXiv:0904.4857 [hep-ph]].

\bibitem{Li:2009xh}
  Y.~Li, F.~Petriello and S.~Quackenbush,
  Phys.\ Rev.\ D {\bf 80} (2009) 055018
  [arXiv:0906.4132 [hep-ph]].

\bibitem{Diener:2009vq}
  R.~Diener, S.~Godfrey and T.~A.~W.~Martin,
  arXiv:0910.1334 [hep-ph].

\bibitem{delAguila:2010mx}
  F.~del Aguila, J.~de Blas and M.~Perez-Victoria,
  JHEP {\bf 1009} (2010) 033
  [arXiv:1005.3998 [hep-ph]].

\bibitem{Chang:2011be}
  C.~-F.~Chang, K.~Cheung and T.~-C.~Yuan,
  JHEP {\bf 1109} (2011) 058
  [arXiv:1107.1133 [hep-ph]].

\bibitem{Erler:2011ud}
  J.~Erler, P.~Langacker, S.~Munir and E.~Rojas,
  JHEP {\bf 1111} (2011) 076
  [arXiv:1103.2659 [hep-ph]].

\bibitem{Chiang:2011kq}
  C.~-W.~Chiang, N.~D.~Christensen, G.~-J.~Ding and T.~Han,
  Phys.\ Rev.\ D {\bf 85} (2012) 015023
  [arXiv:1107.5830 [hep-ph]].

\bibitem{Gopalakrishna:2010xm}
  S.~Gopalakrishna, T.~Han, I.~Lewis, Z.~-g.~Si and Y.~-F.~Zhou,
  Phys.\ Rev.\ D {\bf 82} (2010) 115020
  [arXiv:1008.3508 [hep-ph]].

\bibitem{Frank:2010cj}
  M.~Frank, A.~Hayreter and I.~Turan,
  Phys.\ Rev.\ D {\bf 83} (2011) 035001
  [arXiv:1010.5809 [hep-ph]].

  \bibitem{Schmaltz:2010xr}
  M.~Schmaltz and C.~Spethmann,
  JHEP {\bf 1107} (2011) 046
  [arXiv:1011.5918 [hep-ph]].

\bibitem{Hsieh:2010zr}
  K.~Hsieh, K.~Schmitz, J.~-H.~Yu and C.~-P.~Yuan,
  Phys.\ Rev.\ D {\bf 82} (2010) 035011
  [arXiv:1003.3482 [hep-ph]].

\bibitem{Frank:2010qv}
  M.~Frank, A.~Hayreter and I.~Turan,
  Phys.\ Rev.\ D {\bf 82} (2010) 033012
  [arXiv:1005.3074 [hep-ph]].

\bibitem{Cao:2012ng}
  Q.~-H.~Cao, Z.~Li, J.~-H.~Yu and C.~P.~Yuan,
  Phys.\ Rev.\ D {\bf 86} (2012) 095010
  [arXiv:1205.3769 [hep-ph]].

\bibitem{Du:2012vh}
  C.~Du, H.~-J.~He, Y.~-P.~Kuang, B.~Zhang, N.~D.~Christensen, R.~S.~Chivukula and E.~H.~Simmons,
  Phys.\ Rev.\ D {\bf 86} (2012) 095011
  [arXiv:1206.6022 [hep-ph]].

\bibitem{Duffty:2012rf}
  D.~Duffty and Z.~Sullivan,
  Phys.\ Rev.\ D {\bf 86} (2012) 075018
  [arXiv:1208.4858 [hep-ph]].

\bibitem{CMS12060}
 The CMS Collaboration, CMS-PAS-EXO-12-060.  

  \bibitem{Aad:2012dm}
  G.~Aad {\it et al.}  [ATLAS Collaboration],
  Eur.\ Phys.\ J.\ C {\bf 72} (2012) 2241
  [arXiv:1209.4446 [hep-ex]].

\bibitem{ATLAS:2012ak}
  G.~Aad {\it et al.}  [ATLAS Collaboration],
  Eur.\ Phys.\ J.\ C {\bf 72} (2012) 2056
  [arXiv:1203.5420 [hep-ex]].

  \bibitem{CMS12017}
 The CMS Collaboration, CMS-PAS-EXO-12-017.  

 \bibitem{Chatrchyan:2012gqa}
  S.~Chatrchyan {\it et al.}  [CMS Collaboration],
  Phys.\ Lett.\ B {\bf 718} (2013) 1229
  [arXiv:1208.0956 [hep-ex]].

 \bibitem{Atlas2013050}
 The ATLAS Collaboration, ATLAS-CONF-2013-050.     

\bibitem{Godfrey:2000hc}
  S.~Godfrey, P.~Kalyniak, B.~Kamal and A.~Leike,
  Phys.\ Rev.\ D {\bf 61} (2000) 113009
  [hep-ph/0001074].


\end{thebibliography}
\end{document}
